\documentstyle[12pt,aps,preprint,epsfig,floats,citesort,rotating]{revtex}

\renewcommand\floatpagefraction{0.9}
\renewcommand {\baselinestretch}{1.}
\def\begineq{\begin{equation}}
\def\endeq{\end{equation}}
\def\be{\begin{equation}}
\def\ee{\end{equation}}
\def\la{\left<}
\def\ra{\right>}

\def\PsfigVersion{1.9}
\ifx\undefined\psfig\else \fi

\let\LaTeXAtSign=\@
\let\@=\relax
\edef\psfigRestoreAt{\catcode`\@=\number\catcode`@\relax}
\catcode`\@=11\relax
\newwrite\@unused
\def\ps@typeout#1{{\let\protect\string\immediate\write\@unused{#1}}}
\ps@typeout{psfig/tex \PsfigVersion}

\def\figurepath{./}

\def\@nnil{\@nil}
\def\@empty{}
\def\@psdonoop#1\@@#2#3{}
\def\@psdo#1:=#2\do#3{\edef\@psdotmp{#2}\ifx\@psdotmp\@empty \else
    \expandafter\@psdoloop#2,\@nil,\@nil\@@#1{#3}\fi}
\def\@psdoloop#1,#2,#3\@@#4#5{\def#4{#1}\ifx #4\@nnil \else
       #5\def#4{#2}\ifx #4\@nnil \else#5\@ipsdoloop #3\@@#4{#5}\fi\fi}
\def\@ipsdoloop#1,#2\@@#3#4{\def#3{#1}\ifx #3\@nnil 
       \let\@nextwhile=\@psdonoop \else
      #4\relax\let\@nextwhile=\@ipsdoloop\fi\@nextwhile#2\@@#3{#4}}
\def\@tpsdo#1:=#2\do#3{\xdef\@psdotmp{#2}\ifx\@psdotmp\@empty \else
    \@tpsdoloop#2\@nil\@nil\@@#1{#3}\fi}
\def\@tpsdoloop#1#2\@@#3#4{\def#3{#1}\ifx #3\@nnil 
       \let\@nextwhile=\@psdonoop \else
      #4\relax\let\@nextwhile=\@tpsdoloop\fi\@nextwhile#2\@@#3{#4}}
\ifx\undefined\fbox
\newdimen\fboxrule
\newdimen\fboxsep
\newdimen\ps@tempdima
\newbox\ps@tempboxa
\long\def\fbox#1{\leavevmode\setbox\ps@tempboxa\hbox{#1}\ps@tempdima\fboxrule
    \advance\ps@tempdima \fboxsep \advance\ps@tempdima \dp\ps@tempboxa
   \hbox{\lower \ps@tempdima\hbox
  {\vbox{\hrule height \fboxrule
          \hbox{\vrule width \fboxrule \hskip\fboxsep
          \vbox{\vskip\fboxsep \box\ps@tempboxa\vskip\fboxsep}\hskip 
                 \fboxsep\vrule width \fboxrule}
                 \hrule height \fboxrule}}}}
\fi
\newread\ps@stream
\newif\ifnot@eof       
\newif\if@noisy        
\newif\if@atend        
\newif\if@psfile       
{\catcode`\%=12\global\gdef\epsf@start{
\def\epsf@PS{PS}
\def\epsf@getbb#1{%
\openin\ps@stream=#1
\ifeof\ps@stream\ps@typeout{Error, File #1 not found}\else
   {\not@eoftrue \chardef\other=12
    \def\do##1{\catcode`##1=\other}\dospecials \catcode`\ =10
    \loop
       \if@psfile
	  \read\ps@stream to \epsf@fileline
       \else{
	  \obeyspaces
          \read\ps@stream to \epsf@tmp\global\let\epsf@fileline\epsf@tmp}
       \fi
       \ifeof\ps@stream\not@eoffalse\else
       \if@psfile\else
       \expandafter\epsf@test\epsf@fileline:. \\%
       \fi
          \expandafter\epsf@aux\epsf@fileline:. \\%
       \fi
   \ifnot@eof\repeat
   }\closein\ps@stream\fi}%
\long\def\epsf@test#1#2#3:#4\\{\def\epsf@testit{#1#2}
			\ifx\epsf@testit\epsf@start\else
\ps@typeout{Warning! File does not start with `\epsf@start'.  It may not be a PostScript file.}
			\fi
			\@psfiletrue} 
{\catcode`\%=12\global\let\epsf@percent=
\long\def\epsf@aux#1#2:#3\\{\ifx#1\epsf@percent
   \def\epsf@testit{#2}\ifx\epsf@testit\epsf@bblit
	\@atendfalse
        \epsf@atend #3 . \\%
	\if@atend	
	   \if@verbose{
		\ps@typeout{psfig: found `(atend)'; continuing search}
	   }\fi
        \else
        \epsf@grab #3 . . . \\%
        \not@eoffalse
        \global\no@bbfalse
        \fi
   \fi\fi}%
\def\epsf@grab #1 #2 #3 #4 #5\\{%
   \global\def\epsf@llx{#1}\ifx\epsf@llx\empty
      \epsf@grab #2 #3 #4 #5 .\\\else
   \global\def\epsf@lly{#2}%
   \global\def\epsf@urx{#3}\global\def\epsf@ury{#4}\fi}%
\def\epsf@atendlit{(atend)} 
\def\epsf@atend #1 #2 #3\\{%
   \def\epsf@tmp{#1}\ifx\epsf@tmp\empty
      \epsf@atend #2 #3 .\\\else
   \ifx\epsf@tmp\epsf@atendlit\@atendtrue\fi\fi}

\chardef\psletter = 11 
\chardef\other = 12

\newif \ifdebug 
\newif\ifc@mpute 
\c@mputetrue 

\let\then = \relax
\def\r@dian{pt }
\let\r@dians = \r@dian
\let\dimensionless@nit = \r@dian
\let\dimensionless@nits = \dimensionless@nit
\def\internal@nit{sp }
\let\internal@nits = \internal@nit
\newif\ifstillc@nverging
\def \Mess@ge #1{\ifdebug \then \message {#1} \fi}

{ 
	\catcode `\@ = \psletter
	\gdef \nodimen {\expandafter \n@dimen \the \dimen}
	\gdef \term #1 #2 #3%
	       {\edef \t@ {\the #1}
		\edef \t@@ {\expandafter \n@dimen \the #2\r@dian}%
		\t@rm {\t@} {\t@@} {#3}%
	       }
	\gdef \t@rm #1 #2 #3%
	       {{%
		\count 0 = 0
		\dimen 0 = 1 \dimensionless@nit
		\dimen 2 = #2\relax
		\Mess@ge {Calculating term #1 of \nodimen 2}%
		\loop
		\ifnum	\count 0 < #1
		\then	\advance \count 0 by 1
			\Mess@ge {Iteration \the \count 0 \space}%
			\Multiply \dimen 0 by {\dimen 2}%
			\Mess@ge {After multiplication, term = \nodimen 0}%
			\Divide \dimen 0 by {\count 0}%
			\Mess@ge {After division, term = \nodimen 0}%
		\repeat
		\Mess@ge {Final value for term #1 of 
				\nodimen 2 \space is \nodimen 0}%
		\xdef \Term {#3 = \nodimen 0 \r@dians}%
		\aftergroup \Term
	       }}
	\catcode `\p = \other
	\catcode `\t = \other
	\gdef \n@dimen #1pt{#1} 
}

\def \Divide #1by #2{\divide #1 by #2} 

\def \Multiply #1by #2
       {{
	\count 0 = #1\relax
	\count 2 = #2\relax
	\count 4 = 65536
	\Mess@ge {Before scaling, count 0 = \the \count 0 \space and
			count 2 = \the \count 2}%
	\ifnum	\count 0 > 32767 
	\then	\divide \count 0 by 4
		\divide \count 4 by 4
	\else	\ifnum	\count 0 < -32767
		\then	\divide \count 0 by 4
			\divide \count 4 by 4
		\else
		\fi
	\fi
	\ifnum	\count 2 > 32767 
	\then	\divide \count 2 by 4
		\divide \count 4 by 4
	\else	\ifnum	\count 2 < -32767
		\then	\divide \count 2 by 4
			\divide \count 4 by 4
		\else
		\fi
	\fi
	\multiply \count 0 by \count 2
	\divide \count 0 by \count 4
	\xdef \product {#1 = \the \count 0 \internal@nits}%
	\aftergroup \product
       }}

\def\r@duce{\ifdim\dimen0 > 90\r@dian \then   
		\multiply\dimen0 by -1
		\advance\dimen0 by 180\r@dian
		\r@duce
	    \else \ifdim\dimen0 < -90\r@dian \then  
		\advance\dimen0 by 360\r@dian
		\r@duce
		\fi
	    \fi}

\def\Sine#1%
       {{%
	\dimen 0 = #1 \r@dian
	\r@duce
	\ifdim\dimen0 = -90\r@dian \then
	   \dimen4 = -1\r@dian
	   \c@mputefalse
	\fi
	\ifdim\dimen0 = 90\r@dian \then
	   \dimen4 = 1\r@dian
	   \c@mputefalse
	\fi
	\ifdim\dimen0 = 0\r@dian \then
	   \dimen4 = 0\r@dian
	   \c@mputefalse
	\fi
	\ifc@mpute \then
		\divide\dimen0 by 180
		\dimen0=3.141592654\dimen0
		\dimen 2 = 3.1415926535897963\r@dian 
		\divide\dimen 2 by 2 
		\Mess@ge {Sin: calculating Sin of \nodimen 0}%
		\count 0 = 1 
		\dimen 2 = 1 \r@dian 
		\dimen 4 = 0 \r@dian 
		\loop
			\ifnum	\dimen 2 = 0 
			\then	\stillc@nvergingfalse 
			\else	\stillc@nvergingtrue
			\fi
			\ifstillc@nverging 
			\then	\term {\count 0} {\dimen 0} {\dimen 2}%
				\advance \count 0 by 2
				\count 2 = \count 0
				\divide \count 2 by 2
				\ifodd	\count 2 
				\then	\advance \dimen 4 by \dimen 2
				\else	\advance \dimen 4 by -\dimen 2
				\fi
		\repeat
	\fi		
			\xdef \sine {\nodimen 4}%
       }}

\def\Cosine#1{\ifx\sine\UnDefined\edef\Savesine{\relax}\else
		             \edef\Savesine{\sine}\fi
	{\dimen0=#1\r@dian\advance\dimen0 by 90\r@dian
	 \Sine{\nodimen 0}
	 \xdef\cosine{\sine}
	 \xdef\sine{\Savesine}}}	      

\def\psdraft{
	\def\@psdraft{0}
}
\def\psfull{
	\def\@psdraft{100}
}

\psfull

\newif\if@scalefirst
\def\psscalefirst{\@scalefirsttrue}
\def\psrotatefirst{\@scalefirstfalse}
\psrotatefirst

\newif\if@draftbox
\def\psnodraftbox{
	\@draftboxfalse
}
\def\psdraftbox{
	\@draftboxtrue
}
\@draftboxtrue

\newif\if@prologfile
\newif\if@postlogfile
\def\pssilent{
	\@noisyfalse
}
\def\psnoisy{
	\@noisytrue
}
\psnoisy
\newif\if@bbllx
\newif\if@bblly
\newif\if@bburx
\newif\if@bbury
\newif\if@height
\newif\if@width
\newif\if@rheight
\newif\if@rwidth
\newif\if@angle
\newif\if@clip
\newif\if@verbose
\def\@p@@sclip#1{\@cliptrue}

\newif\if@decmpr

\def\@p@@sfigure#1{\def\@p@sfile{null}\def\@p@sbbfile{null}
	        \openin1=#1.bb
		\ifeof1\closein1
	        	\openin1=\figurepath#1.bb
			\ifeof1\closein1
			        \openin1=#1
				\ifeof1\closein1%
				       \openin1=\figurepath#1
					\ifeof1
					   \ps@typeout{Error, File #1 not found}
						\if@bbllx\if@bblly
				   		\if@bburx\if@bbury
			      				\def\@p@sfile{#1}%
			      				\def\@p@sbbfile{#1}%
							\@decmprfalse
				  	   	\fi\fi\fi\fi
					\else\closein1
				    		\def\@p@sfile{\figurepath#1}%
				    		\def\@p@sbbfile{\figurepath#1}%
						\@decmprfalse
	                       		\fi%
			 	\else\closein1%
					\def\@p@sfile{#1}
					\def\@p@sbbfile{#1}
					\@decmprfalse
			 	\fi
			\else
				\def\@p@sfile{\figurepath#1}
				\def\@p@sbbfile{\figurepath#1.bb}
				\@decmprtrue
			\fi
		\else
			\def\@p@sfile{#1}
			\def\@p@sbbfile{#1.bb}
			\@decmprtrue
		\fi}

\def\@p@@sfile#1{\@p@@sfigure{#1}}

\def\@p@@sbbllx#1{
		\@bbllxtrue
		\dimen100=#1
		\edef\@p@sbbllx{\number\dimen100}
}
\def\@p@@sbblly#1{
		\@bbllytrue
		\dimen100=#1
		\edef\@p@sbblly{\number\dimen100}
}
\def\@p@@sbburx#1{
		\@bburxtrue
		\dimen100=#1
		\edef\@p@sbburx{\number\dimen100}
}
\def\@p@@sbbury#1{
		\@bburytrue
		\dimen100=#1
		\edef\@p@sbbury{\number\dimen100}
}
\def\@p@@sheight#1{
		\@heighttrue
		\dimen100=#1
   		\edef\@p@sheight{\number\dimen100}
}
\def\@p@@swidth#1{
		\@widthtrue
		\dimen100=#1
		\edef\@p@swidth{\number\dimen100}
}
\def\@p@@srheight#1{
		\@rheighttrue
		\dimen100=#1
		\edef\@p@srheight{\number\dimen100}
}
\def\@p@@srwidth#1{
		\@rwidthtrue
		\dimen100=#1
		\edef\@p@srwidth{\number\dimen100}
}
\def\@p@@sangle#1{
		\@angletrue
		\edef\@p@sangle{#1} 
}
\def\@p@@ssilent#1{ 
		\@verbosefalse
}
\def\@p@@sprolog#1{\@prologfiletrue\def\@prologfileval{#1}}
\def\@p@@spostlog#1{\@postlogfiletrue\def\@postlogfileval{#1}}
\def\@cs@name#1{\csname #1\endcsname}
\def\@setparms#1=#2,{\@cs@name{@p@@s#1}{#2}}
\def\ps@init@parms{
		\@bbllxfalse \@bbllyfalse
		\@bburxfalse \@bburyfalse
		\@heightfalse \@widthfalse
		\@rheightfalse \@rwidthfalse
		\def\@p@sbbllx{}\def\@p@sbblly{}
		\def\@p@sbburx{}\def\@p@sbbury{}
		\def\@p@sheight{}\def\@p@swidth{}
		\def\@p@srheight{}\def\@p@srwidth{}
		\def\@p@sangle{0}
		\def\@p@sfile{} \def\@p@sbbfile{}
		\def\@p@scost{10}
		\def\@sc{}
		\@prologfilefalse
		\@postlogfilefalse
		\@clipfalse
		\if@noisy
			\@verbosetrue
		\else
			\@verbosefalse
		\fi
}
\def\parse@ps@parms#1{
	 	\@psdo\@psfiga:=#1\do
		   {\expandafter\@setparms\@psfiga,}}
\newif\ifno@bb
\def\bb@missing{
	\if@verbose{
		\ps@typeout{psfig: searching \@p@sbbfile \space  for bounding box}
	}\fi
	\no@bbtrue
	\epsf@getbb{\@p@sbbfile}
        \ifno@bb \else \bb@cull\epsf@llx\epsf@lly\epsf@urx\epsf@ury\fi
}	
\def\bb@cull#1#2#3#4{
	\dimen100=#1 bp\edef\@p@sbbllx{\number\dimen100}
	\dimen100=#2 bp\edef\@p@sbblly{\number\dimen100}
	\dimen100=#3 bp\edef\@p@sbburx{\number\dimen100}
	\dimen100=#4 bp\edef\@p@sbbury{\number\dimen100}
	\no@bbfalse
}
\newdimen\p@intvaluex
\newdimen\p@intvaluey
\def\rotate@#1#2{{\dimen0=#1 sp\dimen1=#2 sp
		  \global\p@intvaluex=\cosine\dimen0
		  \dimen3=\sine\dimen1
		  \global\advance\p@intvaluex by -\dimen3
		  \global\p@intvaluey=\sine\dimen0
		  \dimen3=\cosine\dimen1
		  \global\advance\p@intvaluey by \dimen3
		  }}
\def\compute@bb{
		\no@bbfalse
		\if@bbllx \else \no@bbtrue \fi
		\if@bblly \else \no@bbtrue \fi
		\if@bburx \else \no@bbtrue \fi
		\if@bbury \else \no@bbtrue \fi
		\ifno@bb \bb@missing \fi
		\ifno@bb \ps@typeout{FATAL ERROR: no bb supplied or found}
			\no-bb-error
		\fi
		\count203=\@p@sbburx
		\count204=\@p@sbbury
		\advance\count203 by -\@p@sbbllx
		\advance\count204 by -\@p@sbblly
		\edef\ps@bbw{\number\count203}
		\edef\ps@bbh{\number\count204}
		\if@angle 
			\Sine{\@p@sangle}\Cosine{\@p@sangle}
	        	{\dimen100=\maxdimen\xdef\r@p@sbbllx{\number\dimen100}
					    \xdef\r@p@sbblly{\number\dimen100}
			                    \xdef\r@p@sbburx{-\number\dimen100}
					    \xdef\r@p@sbbury{-\number\dimen100}}
                        \def\minmaxtest{
			   \ifnum\number\p@intvaluex<\r@p@sbbllx
			      \xdef\r@p@sbbllx{\number\p@intvaluex}\fi
			   \ifnum\number\p@intvaluex>\r@p@sbburx
			      \xdef\r@p@sbburx{\number\p@intvaluex}\fi
			   \ifnum\number\p@intvaluey<\r@p@sbblly
			      \xdef\r@p@sbblly{\number\p@intvaluey}\fi
			   \ifnum\number\p@intvaluey>\r@p@sbbury
			      \xdef\r@p@sbbury{\number\p@intvaluey}\fi
			   }
			\rotate@{\@p@sbbllx}{\@p@sbblly}
			\minmaxtest
			\rotate@{\@p@sbbllx}{\@p@sbbury}
			\minmaxtest
			\rotate@{\@p@sbburx}{\@p@sbblly}
			\minmaxtest
			\rotate@{\@p@sbburx}{\@p@sbbury}
			\minmaxtest
			\edef\@p@sbbllx{\r@p@sbbllx}\edef\@p@sbblly{\r@p@sbblly}
			\edef\@p@sbburx{\r@p@sbburx}\edef\@p@sbbury{\r@p@sbbury}
		\fi
		\count203=\@p@sbburx
		\count204=\@p@sbbury
		\advance\count203 by -\@p@sbbllx
		\advance\count204 by -\@p@sbblly
		\edef\@bbw{\number\count203}
		\edef\@bbh{\number\count204}
}
\def\in@hundreds#1#2#3{\count240=#2 \count241=#3
		     \count100=\count240	
		     \divide\count100 by \count241
		     \count101=\count100
		     \multiply\count101 by \count241
		     \advance\count240 by -\count101
		     \multiply\count240 by 10
		     \count101=\count240	
		     \divide\count101 by \count241
		     \count102=\count101
		     \multiply\count102 by \count241
		     \advance\count240 by -\count102
		     \multiply\count240 by 10
		     \count102=\count240	
		     \divide\count102 by \count241
		     \count200=#1\count205=0
		     \count201=\count200
			\multiply\count201 by \count100
		 	\advance\count205 by \count201
		     \count201=\count200
			\divide\count201 by 10
			\multiply\count201 by \count101
			\advance\count205 by \count201
		     \count201=\count200
			\divide\count201 by 100
			\multiply\count201 by \count102
			\advance\count205 by \count201
		     \edef\@result{\number\count205}
}
\def\compute@wfromh{
		\in@hundreds{\@p@sheight}{\@bbw}{\@bbh}
		\edef\@p@swidth{\@result}
}
\def\compute@hfromw{
	        \in@hundreds{\@p@swidth}{\@bbh}{\@bbw}
		\edef\@p@sheight{\@result}
}
\def\compute@handw{
		\if@height 
			\if@width
			\else
				\compute@wfromh
			\fi
		\else 
			\if@width
				\compute@hfromw
			\else
				\edef\@p@sheight{\@bbh}
				\edef\@p@swidth{\@bbw}
			\fi
		\fi
}
\def\compute@resv{
		\if@rheight \else \edef\@p@srheight{\@p@sheight} \fi
		\if@rwidth \else \edef\@p@srwidth{\@p@swidth} \fi
}
\def\compute@sizes{
	\compute@bb
	\if@scalefirst\if@angle
	\if@width
	   \in@hundreds{\@p@swidth}{\@bbw}{\ps@bbw}
	   \edef\@p@swidth{\@result}
	\fi
	\if@height
	   \in@hundreds{\@p@sheight}{\@bbh}{\ps@bbh}
	   \edef\@p@sheight{\@result}
	\fi
	\fi\fi
	\compute@handw
	\compute@resv}

\def\psfig#1{\vbox {
	\ps@init@parms
	\parse@ps@parms{#1}
	\compute@sizes
	\ifnum\@p@scost<\@psdraft{
		\special{ps::[begin] 	\@p@swidth \space \@p@sheight \space
				\@p@sbbllx \space \@p@sbblly \space
				\@p@sbburx \space \@p@sbbury \space
				startTexFig \space }
		\if@angle
			\special {ps:: \@p@sangle \space rotate \space} 
		\fi
		\if@clip{
			\if@verbose{
				\ps@typeout{(clip)}
			}\fi
			\special{ps:: doclip \space }
		}\fi
		\if@prologfile
		    \special{ps: plotfile \@prologfileval \space } \fi
		\if@decmpr{
			\if@verbose{
				\ps@typeout{psfig: including \@p@sfile.Z \space }
			}\fi
			\special{ps: plotfile "`zcat \@p@sfile.Z" \space }
		}\else{
			\if@verbose{
				\ps@typeout{psfig: including \@p@sfile \space }
			}\fi
			\special{ps: plotfile \@p@sfile \space }
		}\fi
		\if@postlogfile
		    \special{ps: plotfile \@postlogfileval \space } \fi
		\special{ps::[end] endTexFig \space }
		\vbox to \@p@srheight sp{
			\hbox to \@p@srwidth sp{
				\hss
			}
		\vss
		}
	}\else{
		\if@draftbox{		
			\hbox{\frame{\vbox to \@p@srheight sp{
			\vss
			\hbox to \@p@srwidth sp{ \hss \@p@sfile \hss }
			\vss
			}}}
		}\else{
			\vbox to \@p@srheight sp{
			\vss
			\hbox to \@p@srwidth sp{\hss}
			\vss
			}
		}\fi

	}\fi
}}
\psfigRestoreAt
\let\@=\LaTeXAtSign

\renewcommand{\topfraction}{0.9}                                                                                                                                                    
\renewcommand{\bottomfraction}{0.9}                                                                                                                                                 
\renewcommand{\textfraction}{0.05}                                                                                                                                                  
\renewcommand{\floatpagefraction}{0.05}                                                                                                                                             
\setcounter{topnumber}{5}                                                                                                                                                           
\setcounter{bottomnumber}{5}                                                                                                                                                        
\setcounter{totalnumber}{10}

\begin{document}
\bibliographystyle{prsty}
\title{Inert gas accumulation in sonoluminescing bubbles}
\author{
Detlef Lohse and 
Sascha Hilgenfeldt 
}
\address{
Fachbereich Physik der Universit\"at Marburg,\\
Renthof 6, 35032 Marburg, Germany 
}

\vspace{1cm}

\date{J. Chem. Physics, in press (received March 1997)}

\maketitle

\begin{abstract}
In this paper we elaborate on the idea
[Lohse et al.,  Phys.\ Rev.\ Lett.\ 78, 1359-1362 (1997)]
that (single) sonoluminescing air bubbles rectify argon.
The reason for the rectification is that nitrogen and oxygen dissociate and
their reaction products dissolve in water. We give further experimental and
theoretical evidence and extend the theory to other gas mixtures.
We show that
in the absence of chemical reactions (e.g., for inert gas mixtures)
gas accumulation in
strongly acoustically driven bubbles can also
occur.
\end{abstract}

\pacs{PACS numbers: 78.60.Mq, 42.65.Re, 43.25.+y, 
47.40.Nm}


\newpage

\section{Introduction}

\begin{figure}[htb]
\setlength{\unitlength}{1cm}
\begin{center}
\begin{picture}(11,9)
\put(-1.0,0.0){\psfig{figure=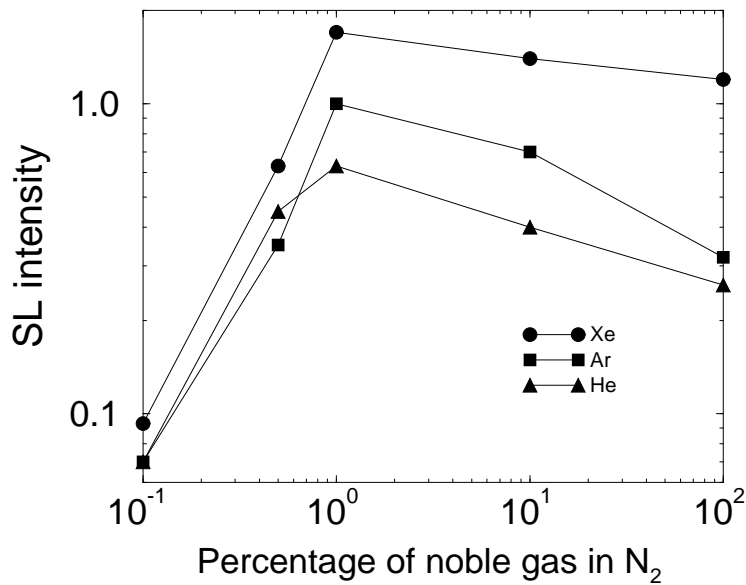,width=12cm,angle=-90}}
\end{picture}
\end{center}
\caption[SL intensity as a function of the nitrogen percentage]{
SL intensity (normalized to air) from a SL bubble in water as a 
function
of the percentage (mole fraction) of noble gas mixed 
with nitrogen.
The gas mixture was dissolved in water at a pressure 
head of 150\,mmHg, i.e., $p_\infty /P_0 \sim 0.2$.
The data are taken from figure 1 of
Hiller et al.\ \cite{hil94}.
}\label{fig8}
\end{figure}

Sonoluminescence (SL) has long been known to be very sensitive to the used
gas \cite{fre34,har39,kut62,mar85,bre95b}.
This effect is even more pronounced
for single bubble sonoluminescence (SBSL),
a phenomenon in which a single gas bubble is driven by a strong acoustic field
and can emit short light pulses for hours \cite{gai90,cru94}.
Detailed experiments by the Putterman group at UCLA
\cite{bar91,bar92,hil92,hil94,hil95,bar95,loe93,loe95,wen95}
revealed that inert gas as part
of the operating gas is essential. In figure
\ref{fig8}
we show the experimental SL intensity from a SL bubble in water as a function of
the percentage $\xi_l$ of noble gas mixed with nitrogen, taken from ref.\
\cite{hil94}. Pure nitrogen bubbles show hardly any SL emission;
the optimum is around
$\xi_l = 0.01 = 1\%$, i.e., the amount of argon contained in air.

Besides the noble gas percentage $\xi_l$ the other experimentally
controllable parameters in SBSL experiments are the forcing pressure amplitude
$P_a$ of the forcing sound field
\be
P(t) = P_a \cos \omega t,
\label{sound}
\ee
the total gas pressure overhead $p_\infty$, and the ambient pressure $P_0$
which
is normally 1atm. The frequency $\omega /2\pi$ which is in a range between 20kHz
and 40kHz is fixed as it has to be adopted to the size of the resonator (``Crum
cell'', \cite{cru75}). Note that the ambient radius $R_0$ of the bubble (i.e.,
the bubble radius at ambient standard conditions $P_0=1$atm and 
at the  temperature of the water)
is {\it not} a free parameter, but the system adjusts
$R_0$ itself by diffusional processes.

There are two types of SBSL: stable and unstable SBSL \cite{bar95,hil96}.
In {\it unstable} SBSL the phase of light emission
(measured relative
to the phase of the forcing sound field) {\it drifts}
on a diffusional timescale of seconds, interrupted
by sudden breakdowns. The same
is true for the maximal bubble size per cycle and for the light intensity. An
example for the bubble dynamics and light intensity for unstable SBSL
is shown in figure
6 of ref.\ \cite{loe95}.
In ref.\ \cite{hil96} we quantitatively accounted for unstable SBSL
as light emission from
a bubble growing by rectified diffusion
\cite{ell69,bre95b,fyr94,loe95,bre96,hol96,hil96}
and thereby running into a shape instability
\cite{ple54,pro77,bre95,hil96,bre95b,hol96}
where it pinches off a microbubble
and starts over. The pinch off leads to a {\it recoil} of the bubble which makes
it appear ``jiggling'' or ``dancing'' \cite{cru94,gai90,bar95}.
Meanwhile, the pinched off microbubble could be visualized \cite{holzfuss}.
The second type
of SBSL is {\it stable}
 SBSL, distinguished by a constant phase and intensity of the
 light pulses, repeating for
hours with remarkable precision \cite{bar92,bar91,hol94}.

In ref.\ \cite{hil96} we calculated 
{\it phase diagrams} of SL bubbles in the ambient radius vs forcing pressure and
gas pressure vs forcing pressure phase spaces.
These diagrams are based on the
Rayleigh-Plesset equation for the bubble radius $R(t)$,
a similar type of approximation for shape distortions,
and the advection diffusion
equation.
We call this approach the
{\it Rayleigh-Plesset SL bubble approach}.
Shape stabilities and diffusional stabilities are considered.
 The phase diagrams 
quantitatively agree with Barber et al.'s \cite{bar95} and L\"ofstedt
et al.'s \cite{loe95} experiments for argon
(and other inert gas) 
bubbles, but {\it not} for air bubbles.

\begin{figure}[htb]
\setlength{\unitlength}{1.0cm}
\begin{center}
\begin{picture}(11,9)
\put(-1.0,0.0){\psfig{figure=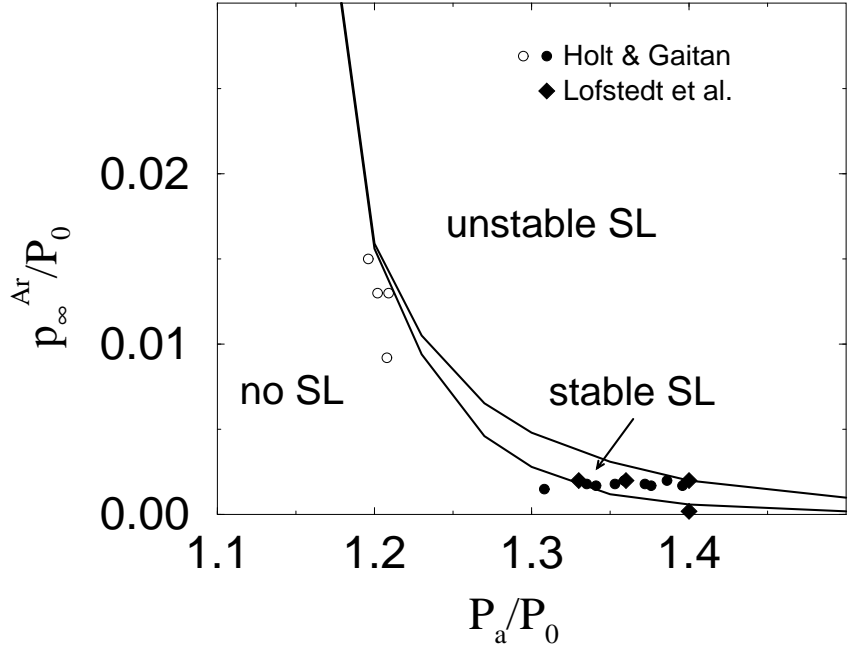,width=12cm,angle=-90}}
\end{picture}
\end{center}
\caption[$p^{Ar}_\infty$-$P_a$ phase diagram with experimental data included]{
Phase diagram for pure argon bubbles in the
$p^{Ar}_\infty/P_0^{}$ versus $P_a/P_0$ parameter space. 
Stable SL is only possible
in a very small window of argon concentration.
The experimental data points included for comparison
refer to observed stable
SL (filled symbols) or stable non-SL bubbles
(open symbols)
from refs.\ \cite{loe95} (diamonds) and \cite{hol96} (circles)
and show good agreement with the theory.
Note that only those data can be included for which
$P_a$, $p_\infty$ and $\xi_l$ are experimentally known.
}
\label{phase_dia_withdata}
\end{figure}

For argon bubbles stable SBSL is only possible in a small window of tiny gas
concentration, see figure \ref{phase_dia_withdata}. For $P_a = 1.3$atm this
window is between $p_\infty^{Ar}/P_0 = 0.002$ and $0.004$
in very good agreement with L\"ofstedt et al.'s experimental data.
 Repeating
the calculation of such phase diagrams for air bubbles essentially gives the
same result, however, experimentally stable SBSL in air bubbles is found at
about one hundred
 times larger gas pressure overhead \cite{bar95,hol96,gai90}. 
This discrepancy between air and argon bubbles was first pointed out by
L\"ofstedt et al.\ \cite{loe95} who
hypothesized 
an ``as yet unidentified
mass ejection mechanism'' in air bubbles which ``is the key to SL in a
single bubble''.

\section{Air bubbles vs argon bubbles}
In refs.\ \cite{bre96b,loh97} we  have suggested
that this mechanism is {\it chemical}.
The occurrence of
chemical reactions has
in fact led to the discovery of 
multibubble sonoluminescence (MBSL) 
\cite{har39,wei53,kut62,mar85,ver88,suslick,bre95b}:
Frenzel and Schultes \cite{fre34}
were stimulated to look for luminescence as the formation of hydrogen peroxide
in aqueous fluids subjected to sound
had been observed before.
Later, Schultes and Gohr \cite{sch36} found that also 
nitric and nitrous acids were produced.
The reason is that 
the high temperatures generated by 
the bubble collapse are beyond the
dissociation temperature of oxygen and nitrogen ($\approx 9000$K
\cite{temperaturefuss}), leading
to the formation of O and N radicals which react with the H and O 
radicals formed from the dissociation of 
water vapor. Rearrangement of the radicals will lead to the formation
of NO, OH, and NH, which eventually dissolve in water to form
H$_2$O$_2$, HNO$_2$, and
HNO$_3$, among other products.

Based on fits of SBSL spectra \cite{ber95,hil92} and hydrodynamic
calculations \cite{loe93,bar94},
it is assumed 
that internal bubble temperatures in SBSL are 
even higher than in MBSL. Therefore, the same reactions as in MBSL
will occur.
The reaction products (NO$_2$, NO, $\dots$) are
pressed into the surrounding liquid, and are not recollected during
the next bubble cycle, since their solubility 
 in water is enormous.
These chemical processes deprive
 the gas in the bubble of its reactive
components.
Small amounts of N$_2$ and O$_2$ that diffuse into
the bubble during the
expansion react and their dissociation products are expelled back into
the surrounding liquids at the bubble collapse.
The only gases that can remain within a SBSL bubble over
many bubble cycles are those which at high temperatures
do not react with the liquid vapor,
i.e., inert gases. 
Hence, when air is dissolved in water, a strongly forced
bubble 
is almost completely 
filled with argon. This  
argon rectification happens in SBSL but not in MBSL because
it requires long time bubble stability.

This argument immediately suggests that 
the {\it partial} pressure
of argon 
\be
p_\infty^{Ar} = \xi_{l} 
p_\infty 
\label{5partial}
\ee
 determines bubble stability, {\it not} the total pressure $p_\infty$.
Indeed, if we include Holt and Gaitan's experimental data for air bubbles 
\cite{hol96} on 
inert gas - nitrogen mixtures and {\it only consider the inert gas
partial pressure} as the relevant quantity for diffusive stability,
excellent agreement with the theoretical phase diagrams
\cite{hil96} is found.

The phase diagram shows that for $P_a = 1.3$atm
argon bubbles exist between $0.002 < 
p_\infty^{Ar}/P_0 < 0.004 $. For {\it pure} argon dissolved in water, 
$p_\infty^{Ar} = p_\infty$.
For air bubbles, however, the partial pressure of argon is only 
$p_\infty^{Ar}=0.01p_\infty$
which requires
$0.2 < p_\infty /P_0 < 0.4$ for stable SL with ``air'' bubbles, 
in good agreement with experiment.  Since this
amount of degassing is easily achieved, air with its $1\%$ argon is 
a particularly friendly gas for SL experiments.
The theory suggests that for an argon ratio of $\xi_{l}
\approx 0.0033$ at
$P_a= 1.3atm$ there is stable SL between $0.6 < p_\infty /P_0 < 1.2$, so that
degassing is not required.
Also, the window of stability is even wider than for air.
The major problem in experimentally achieving SBSL without degassing
is spontaneous cavitation, provoked by impurities
in the liquid.  These must be eliminated for the experiment to work.
Another way to obtain stable SL without degassing would be to slightly
increase the ambient pressure $P_0$ so that the ratio
$p_\infty^{Ar}/P_0$ is in the
required window. 
Indeed, with such a
kind of experiment stable SBSL without degassing could
recently be achieved \cite{kon_priv}.

\begin{figure}[htb]
\setlength{\unitlength}{1.0cm}
\begin{center}
\begin{picture}(11,9)
\put(-0.0,-0.3){\psfig{figure=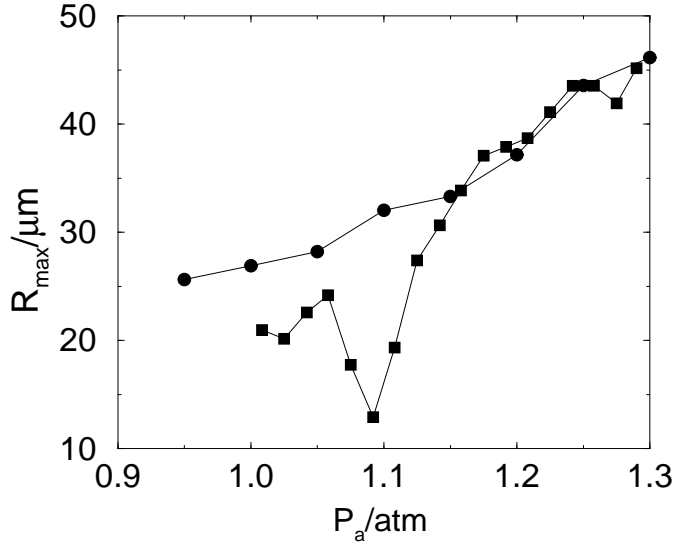,width=11cm,angle=-90}}
\end{picture}
\end{center}
\caption[``Waterfall plots'' for argon and air]{
Transition in the maximal radius
towards the SL regime for argon bubbles (dots) 
and for air bubbles
(squares).
Bubbles around $P_a=1.1$atm start to glow. 
Only for air bubbles a breakdown in the radius is 
seen
near that onset of SL, signaling the threshold for 
nitrogen dissociation.
The data for argon are taken from figure 4 of Hiller et al.'s work
\cite{hil94}, the total gas concentration is about $p_\infty =
150$mmHg.
This means $p^{Ar}_\infty/P_0 = 0.2$ and according to
figure \ref{phase_dia_withdata} the bubble should be in the unstable SL 
regime in
agreement with the observations of ref.\ \cite{hil94}.
The data for air are taken 
from figure 2 of Barber et al.'s work
\cite{bar94}.
The gas saturation is about $10\%$, corresponding to
$p^{ Ar}_\infty/P_0 = 0.01 \cdot 0.1 = 0.001$.
According to figure \ref{phase_dia_withdata} we have stable SL around $P_{a}
\sim 1.4$\,atm
which again is in agreement with the experiment reported in ref.\ \cite{bar94}.
}
\label{argon_and_air}
\end{figure}

\begin{figure}[htb]
\setlength{\unitlength}{1cm}
\begin{center}
\begin{picture}(12,10.)
\put(-0.0,0.0){\psfig{figure=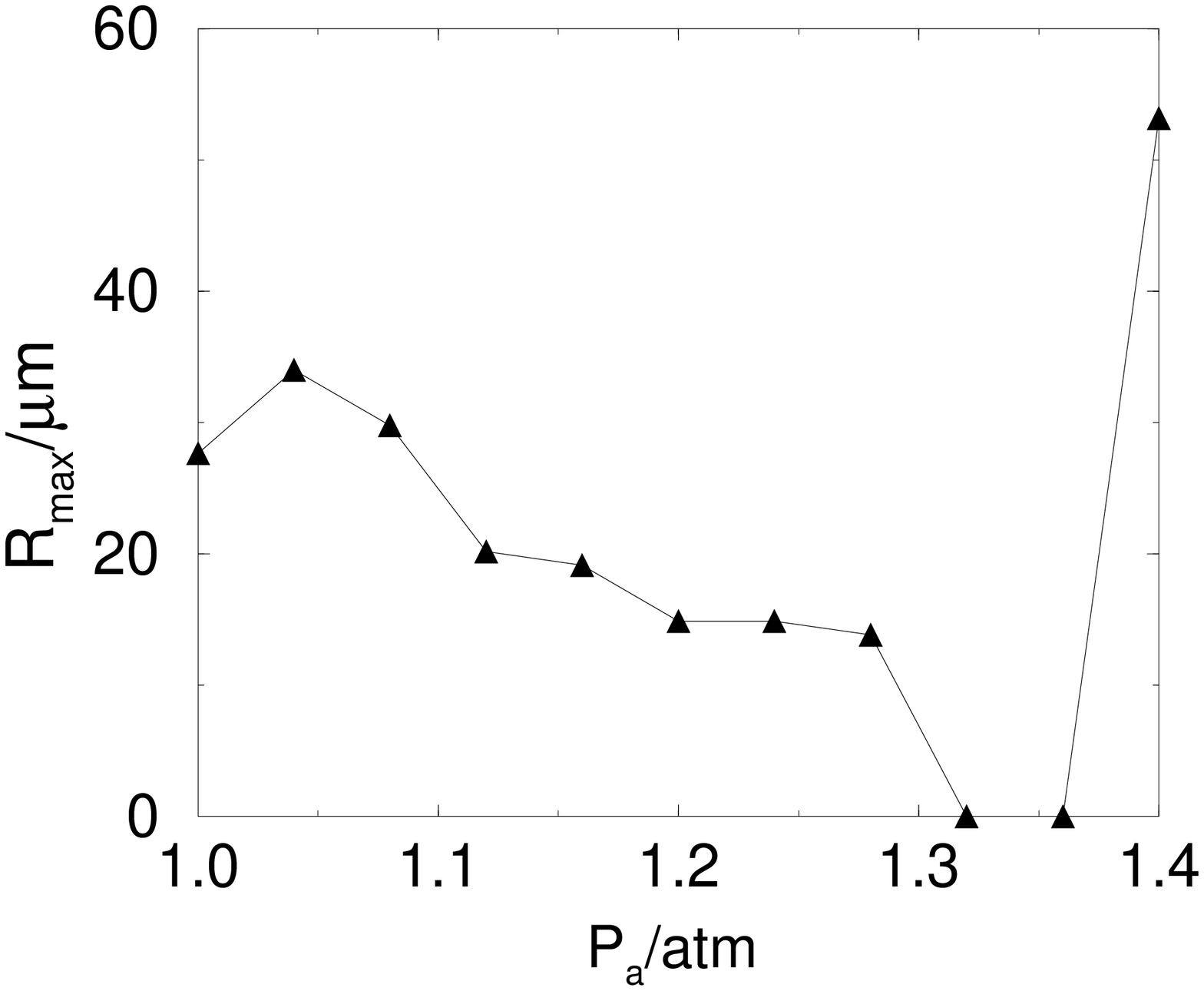,width=11cm,angle=-90}}
\end{picture}
\end{center}
\caption[``Waterfall plot'' for a xenon-nitrogen mixture]{
The transition (in $R_{max})$
to SL for a bubble filled with an initial $0.1\%$
xenon in nitrogen gas mixture at a
partial pressure of 150\,mmHg.
Only the bubble at $P_a=1.4$atm emits light. 
According to the 
dissociation hypothesis this
corresponds to $p^{Xe}_\infty/P_0 = 0.0002$ for a pure sonoluminescing
xenon bubble.
From the  phase diagram figure \ref{phase_dia_withdata} which is (with tiny
corrections) also valid for xenon we conclude
that bubbles driven at $P_{ a}=1.3$\,atm dissolve for these 
low concentrations,
whereas bubbles at $P_{ a}=1.4$\,atm show stable SL, just as 
seen here. The data of this 
figure
are taken from figure 11 of
L\"ofstedt et al.\ \cite{loe95}.
}\label{fig_gap}
\end{figure}

Before we proceed with a quantitative analysis, we will give further
support for the nitrogen dissociation hypothesis
\cite{bre96b,loh97}  from comparison with
various experimental results on gas mixtures. 
\begin{itemize}
\item
{\it Transition towards SL:}
The transition towards SL with increasing forcing pressure $P_{ a}$ is
shown in figure \ref{argon_and_air} for both argon and air bubbles.
For pure argon bubbles the transition to SL is very smooth.
For air, however,
one can observe a breakdown of the bubble radius at about
$1.1$\,atm, signaling that the dissociation threshold of
N$_2$ is achieved. 
Before the transition
the bubble is filled with a 
mixture of nitrogen, oxygen, and argon,
and the 
ambient radius is determined by the combination of all 
three gases.  Beyond 
the dissociation threshold, 
only argon is left in the bubble.
--
The transition from the no SL regime in figure \ref{phase_dia_withdata}
to the SL regime
can also be seen
in figure \ref{fig8}, where the SL light intensity is plotted
as a function of $\xi_l$ for 
fixed $P_{a}$ (we assume
$P_{a}=1.3$atm) and fixed $p_\infty/P_0\sim 0.20$.
According to our theory 
we expect SL for $p_\infty^{Ar}/P_0 > 0.002$ or
\be
\xi_l > {p_\infty^{Ar} / P_0 \over p_\infty/P_0 } = 0.01
\label{5xil_lower}
\ee
in pretty good
agreement with figure \ref{fig8} where we indeed observe that
strong SL is ``switched on'' at about that concentration.
Our theory also predicts that near the switch on we always
have stable SL, whereas for larger $\xi_l$ {\it unstable} SL develops.
The window of stable SL is only in between $0.002 < p_\infty^{Ar} /P_0
< 0.004$, so that beyond $\xi_l = 0.004/p_\infty$ we expect 
unstable
SL, a prediction which should be verified.
\item
{\it Hysteresis:}
Above we have seen that by decreasing the percentage $\xi_l$ of inert gas down
to
$\xi_l\approx 0.0033$
at $P_a=1.3$atm, we obtain a very wide window of stable SL
around $p_\infty/P_0=1$.
When the percentage $\xi_{l}$ of inert gas is even lower, the
argon partial pressure $p_\infty^{Ar}/P_0 = \xi_{l} p_\infty /P_0$
can be pushed below the stable SL regime in figure
\ref{phase_dia_withdata} even for 
$p_\infty / P_0 =1$.
An example of this has been observed  in figure 11 of
L\"ofstedt et al.\ \cite{loe95}, see figure \ref{fig_gap} of the present work,
which studies SBSL of xenon doped nitrogen bubbles (which 
behave like argon doped nitrogen
bubbles) in water at $\xi_{l} =0.001$ and
$p_\infty/P_0=0.2$,
corresponding to $p_\infty^{Ar}/P_0 = 0.0002$,
provided that the gas temperature is high enough to exceed the 
dissociation temperature of nitrogen. This is the case at $P_a = 1.3$atm
and according to figure
\ref{phase_dia_withdata} no stable SL is possible. However, for larger 
$P_a = 1.4$atm stable SL becomes possible again (as the window of stable SL
moves down), just as observed in experiment.
Moreover, the system shows hysteresis:
 the SL state at $P_a=1.4$atm
can be reached by continuously increasing the forcing and thus
continuously
substituting
N$_2$ with xenon.
On the other hand, decreasing the forcing pressure below
the stability threshold for pure xenon bubbles
leads to the dissolution of the bubble.
\item
{\it Unstable SL:}
An example for unstable SL is  shown in figure 6 of ref.\ \cite{loe95}.
For that figure we have $\xi_l = 0.05$ and $p_\infty/P_0 = 0.20$. 
Thus, from  equation
(\ref{5partial}) we have $p_\infty^{ Ar}/P_0 =0.01$ and 
according to the phase diagram figure \ref{phase_dia_withdata}
we are well in the unstable SL regime, just in
agreement with the observations.
\item
{\it Large $P_a$ bubble dissolution islands:}
Further support comes from the recent work of  
Holt and Gaitan \cite{hol96} who measured
detailed phase diagrams as a function of the ambient radius
$R_0$ and the forcing pressure $P_a$ for
different air pressure overhead  $p_\infty/P_0$.
Their central result is that 
at $p_\infty / P_0 = 0.2$ there is a relatively
large forcing pressure $P_a 
\sim 1.2-1.3atm$ regime where bubbles dissolve, see figure \ref{holt}.
Such dissolution
islands do not exist 
within theories of rectified diffusion 
\cite{ell69,fyr94,loe95,hil96}, which predict the bubbles to grow in
that regime.
However,  
considering the nitrogen dissociation at large $P_a$, one realizes
that the observed dissolution islands are a direct consequence:
chemical reactions deplete the bubbles from air and only  argon is left.
As the relevant partial pressure  $p_\infty^{Ar}/P_0 = 0.002$ is 
so low, 
bubbles must indeed   shrink in that regime. 
\item
{\it Isotope scrambling:}
Also experiments with hydrogen gas  support the dissociation hypothesis.
Hiller et al.\ \cite{hil95} analyze SL in 
H$_2$ and
D$_2$ gas bubbles, both in normal and in heavy water. 
Since the gas dynamics inside the bubble determines the
strength of the light emission,
the SL intensity curves should 
group
according to the gas content.
However the four
(H$_2$ in H$_2$O,
 H$_2$ in D$_2$O,
 D$_2$ in H$_2$O, and
 D$_2$ in D$_2$O)
experiments group according to the 
surrounding
liquids, cf.\ figure 2 of \cite{hil95}. 
This suggests the following scenario:
Both the gas and the liquid vapor in the bubble
dissociate to some extent during the (hot) compression phase and 
recombine later on during
expansion. 
Even if there is only a minor amount of this
isotope ``scrambling'', after thousands of cycles the gas
in solution around the bubble would contain the
same hydrogen isotope as the bulk liquid.
This kind of ``scrambling'' is well known in MBSL \cite{har87}. 
The 
acoustic resonator theory of SBSL
\cite{bre96d} stipulates
that the light intensity decreases with increasing acoustic
transmission,
which scales with the ratio of
the gas density to the liquid density.
Since D$_2$ is heavier than H$_2$, SBSL in 
heavy water should be dimmer than in normal
water, as observed by Hiller and Putterman \cite{hil95}.
\end{itemize}

\begin{figure}[t]
\vspace{14cm}
\caption[Experimental phase diagram in $R_0$-$P_a$ space]{
This figure is adopted from Holt and Gaitan's
measurement figure 1c of ref.\ \cite{hol96}.
It is the phase diagram in the $R_0$-$P_a$ parameter space for air at
$p_\infty/P_0=0.20$; the driving frequency is $20.6$kHz.
Arrows indicate whether the bubbles grow or shrink.
Three equilibrium curves A, B, and C can be recognized.
In between curves B and C there is a ``dissolution island''. 
The shaded area shows the shape stable parameter domain.
Note that in the SL region
the onset of shape instabilities is at $R_0 \approx 7\mu m$ rather than
at $R_0 \approx 5\mu m$ as theoretically calculated in ref.\ \cite{hil96}.
One contribution to this deviation is that
the driving frequency chosen by Holt and Gaitan
is smaller than the $\omega/ 2\pi = 26.5$kHz
used by Barber et al.\ \cite{bar94} and throughout the calculations presented
here; another one is the oversimplified model of thermal effects we use. 
}\label{holt}
\end{figure}

\begin{figure}[htb]
\setlength{\unitlength}{1.0cm}
\begin{center}
\begin{picture}(11,9)
\put(-2.0,-0.3){\psfig{figure=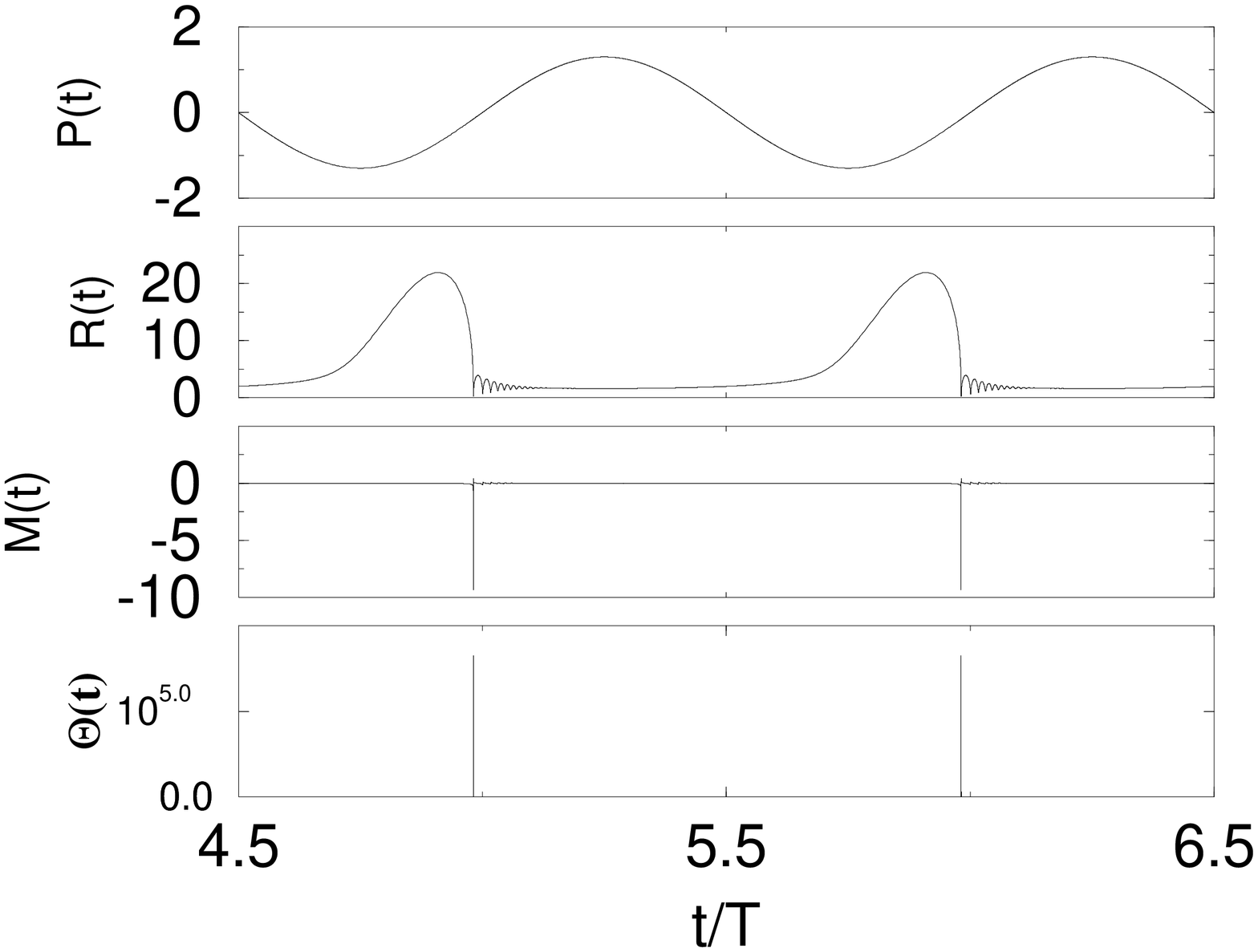,width=13cm,angle=-90}}
\end{picture}
\end{center}
\caption[Mach number and temperature as function of time]{
Forcing pressure (in atm) with amplitude
$P_a=1.3$atm, bubble radius $R(t)$ (in $\mu m$) with the
typical collapse (``Rayleigh collapse'') after the maximum,
Mach number $M(t) = \dot R(t) / c_g(t)$ (where $c_g$ is the speed
of sound in the gas), and temperature $\Theta (t)$ (in $K$) as
function of time for a bubble with $R_0=2\mu m$ (top to bottom).
}
\label{rp_dyn_various}
\end{figure}

\section{Modeling thermal effects within Rayleigh-Plesset bubble dynamics}
We now proceed to a quantitative calculation of phase diagrams
for gas mixtures. 
The  dynamics of the bubble radius $R(t)$ 
is well described by the Rayleigh-Plesset equation
\cite{ray17,bre95b}.
\begin{eqnarray}
R \ddot R + {3\over 2} \dot R^2  &=&
{1\over \rho_l} \left(p(R,t) - P(t) - P_0 \right)
             \nonumber \\
	            &+& {R\over \rho_l c_l} {d\over dt}
		    p(R,t)
		    - 4 \nu 
{\dot R \over R} -
		    {2\sigma \over
		    \rho_l R}
		    \label{rp}
\end{eqnarray}
with a
van der Waals pressure
\begineq
p(R(t)) = \left(P_0 + {2\sigma \over R_0}\right)
 \left( {R_0^3 - h^3\over R^3(t) - h^3 
}\right)^\gamma .
\label{pressure}
\endeq
Typical parameters for an argon bubble in water at room temperature
\cite{bar95} are
the surface tension
$\sigma = 0.073 kg/s^2$, the water viscosity 
$\nu = 10^{-6} m^2/s$, density 
$\rho_l= 1000 kg/m^3$,
the polytropic exponent $\gamma =1$,
and
speed of sound in water $c_l=1481 m/s$.
For all calculations in this work
we picked the same  driving frequency as in
the SL experiments performed on argon bubbles
\cite{bar95}, namely  
$\omega/ 2\pi = 26.5 kHz$, corresponding to a period $T=2\pi/\omega = 38\mu s$.
Finally, 
$h= R_0/8.86$
is the hard core van der Waals radius for argon bubbles \cite{loe93}.
Typical time series for the bubble radius $R(t)$
resulting from equation (\ref{rp})  for given forcing
$P(t)$ are shown in fig.\ \ref{rp_dyn_various}.

{\it Thermal } conduction effects
 can approximately be taken into account  by simply putting the polytropic exponent
$\gamma = 1$
in equation (\ref{pressure}).
The reason is that the bubbles employed in
SL experiments are so tiny and the oscillation period
$T\approx 38 \mu s$ so long that the gas in the bubble equilibrates with the
water temperature $\Theta_l$. A quantitative analysis was carried out 
by 
Prosperetti \cite{pro77b}
(see also the review \cite{ple77}) who
calculated
 how the polytropic exponent
$\gamma$  depends on 
the (thermal)
Peclet number
$Pe = R_0^2\omega/\kappa$, see figure 1 of ref.\ \cite{ple77}. 
For experimental support of this analysis, see Crum \cite{cru83}.
The Peclet number 
gives the ratio between 
the bubble length scale $R_0$
and the thermal diffusion length 
$\sqrt{\kappa / \omega}$.
The thermal diffusivity $\kappa$ for argon is
$\kappa\approx 2\cdot10^{-5}m^2/s$,
which yields $Pe\approx 0.2$ for $R_0 = 5\mu m$
and according to figure 1 
of ref.\ \cite{ple77}, the  polytropic exponent $\gamma=1$.

The RP equation obviously
contains much smaller time
scales than $\omega^{-1}$.
One could therefore argue that these smaller 
timescales may enter
into the calculation of $Pe$,  so that
the
frequency $\omega$ should be replaced by $|\dot R|/R$.
This estimate leads to an instantaneous Peclet number 
 \be Pe = {|\dot R |
R_0^2\over 
R \kappa }
\label{2peclet}
\ee 
which can become as large as
as $10^4$ at the Rayleigh collapse.
According to figure 1 of ref.\ \cite{ple77}
showing $\gamma (Pe)$
or to 
Prosperetti, Crum, and Commander's fit of that   curve \cite{pro88} 
this implies 
$\gamma \approx
5/3$ for argon at the time of the collapse.
However, since $Pe (t) \gg 1$ only holds in 
very small time
intervals $\sim 1ns$, the global dynamics  are
hardly affected by setting
the effective polytropic  exponent $\gamma = 1$ uniformly
in time. Note 
that with $\gamma = 1$
equation (\ref{pressure}) should not be thought of as 
an equation of state but rather
as a process
equation parametrizing the isothermal conditions at the 
bubble wall,
induced by the large heat capacity of water.
The choice of $\gamma=1$ is
confirmed by the full numerical simulations of Vuong and 
Szeri \cite{vuo96}
and by the analysis of 
Kamath et al.\ \cite{kam93}. 
Note that, as a 
consequence, there are heat
fluxes back and forth across the bubble wall.

Another confirmation for putting $\gamma = 1$ for the bubble dynamics
comes from a recent work by 
Yasui \cite{yas95}
who modeled the heat flow through the bubble wall and also included
water evaporation and condensation.
Consideration of these effects  
\cite{yas95}
fits the experimentally
measured $R(t)$ curves
\cite{loe93,bar94,bar95} pretty well
\cite{wen97}.
However, the essence of the dynamics is
the same as with simply putting $\gamma = 1$
and the phase diagrams we are
going to calculate in this paper are hardly affected
by neglecting the details of the thermal effects.

Even if thermal effects are simply treated by putting
$\gamma = 1$ for most of the time during
the bubble oscillation, they are essential at the final stage of
the collapse and thus 
for estimates of the {\it temperature} achieved in the bubble. 
The present understanding
\cite{pro88,bre95b} is that the only 
way to reliably predict the temperatures in the bubble
is a full numerical simulation
 of the gas dynamical equations inside the bubble.
Clearly, for strong enough forcing shock waves will bounce back
and forth inside the bubble and 
the spatial temperature distribution in
the bubble will be highly inhomogeneous
\cite{wu93}.

Such full numerical calculation is not within the spirit of the 
 Rayleigh-Plesset  SL bubble approach \cite{hil96}. 
We therefore come back to Prosperetti's  \cite{pro77b}
model and estimate also in the case of nonlinear bubble oscillations
the temperature within that model. I.e., we calculate the instantaneous
Peclet number (\ref{2peclet}),
determine the resulting effective polytropic exponent $\gamma_{\it eff}$
from figure 1 of \cite{ple77} 
and assume a van der Waals behavior of the
internal gas according to
\begin{equation}
\Theta(t) = \Theta_l \left( { R_0^3 -h^3  \over
R^3(t) - h^3 }
\right)^{\gamma_{\it eff} - 1 },
\label{2temp}
\end{equation}
where $\Theta_l$ is the temperature of the liquid. This, 
as any model
which neglects spatial inhomogeneities in the gas, is a very
crude model for the temperature and will only be able to give
orders of magnitude.
The physical statements resulting from it, however, do {\it not} depend
on details of this model and so we feel that such an approach is justified.

\begin{figure}[htb]
\setlength{\unitlength}{1.0cm}
\begin{center}
\begin{picture}(11,9)
\put(-2.0,-0.2){\psfig{figure=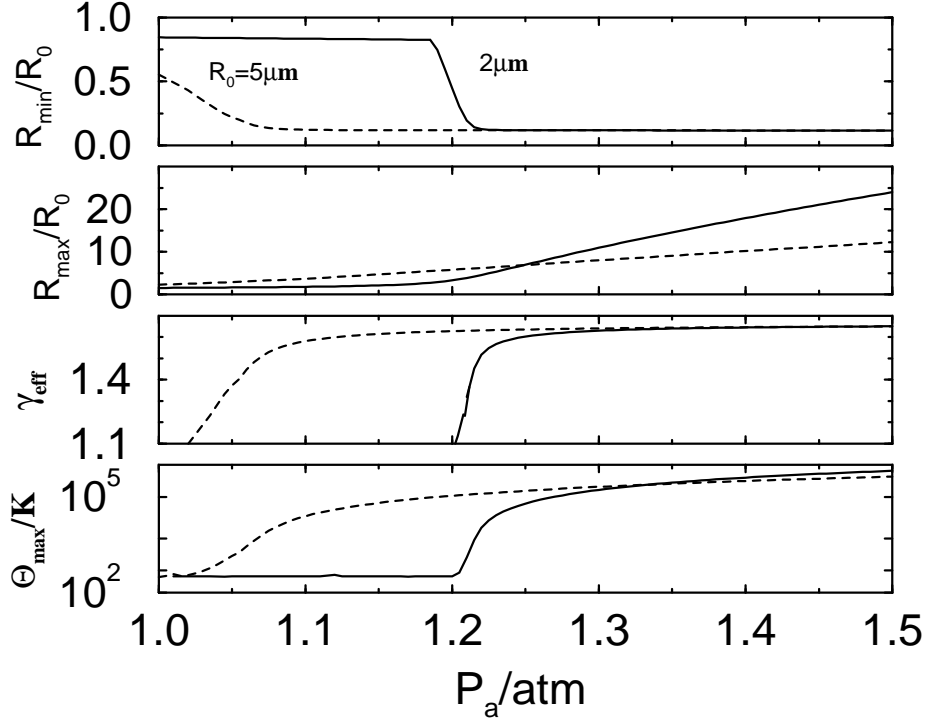,width=13cm,angle=-90}}
\end{picture}
\end{center}
\caption[Crossover in bubble dynamics]{
The minimal radius, the maximal radius, the
maximal effective polytropic exponent
$\gamma_{eff}$, and the maximal temperature $\Theta_{max}$ 
are shown for two different ambient radii
$R_0=2\mu m$ (solid) and $R_0=5\mu m$ (dashed)
as a function of the forcing
pressure amplitude $P_a$. For smaller bubbles the crossover from
the harmonically oscillating regime for small $P_a$ to the bouncing bubble
regime for large $P_a$ is more abrupt.
}
\label{rp_r02and5}
\end{figure}

It can be seen from
figures \ref{rp_r02and5}c and \ref{rp_dyn_various}d
that this temperature model gives the correct trends. 
In figure
\ref{rp_r02and5}c we show the {\it maximal} $\gamma_{eff}$ 
per cycle as a function of $P_a$.
We also give the minimal and the maximal bubble radius in figs.\
\ref{rp_r02and5}a and b. 
For small $P_a$ the maximal effective polytropic exponent
$\gamma_{eff}$  is close to one and jumps
towards $5/3$ at the transition from the sinusoidal to the bouncing bubble
$R(t)$ dynamics.
This transition has been analyzed in detail in ref.\ \cite{hil97}. 
The temperature $\Theta (t)$ from equation (\ref{2temp}) is
shown in figure \ref{rp_dyn_various}d. It strongly peaks at the collapse; for
all other times it essentially 
equals the liquid temperature $\Theta_l $ ($=293$K in this example).
The heating
at the collapse can be enormous.

\begin{figure}[htb]
\setlength{\unitlength}{1.0cm}
\begin{picture}(11,7.0)(0.3,2.2)
\put(12.4,5.3){$\Theta /K$}
   \put(9.8,7.8){
      \begin{rotate}{45}
             $R_0/\mu m$
        \end{rotate}}
    \put(6.9,2.6){
        \begin{rotate}{-7}
             $P_a/atm$
        \end{rotate}}
\put(0.5,-3.3){\psfig{figure=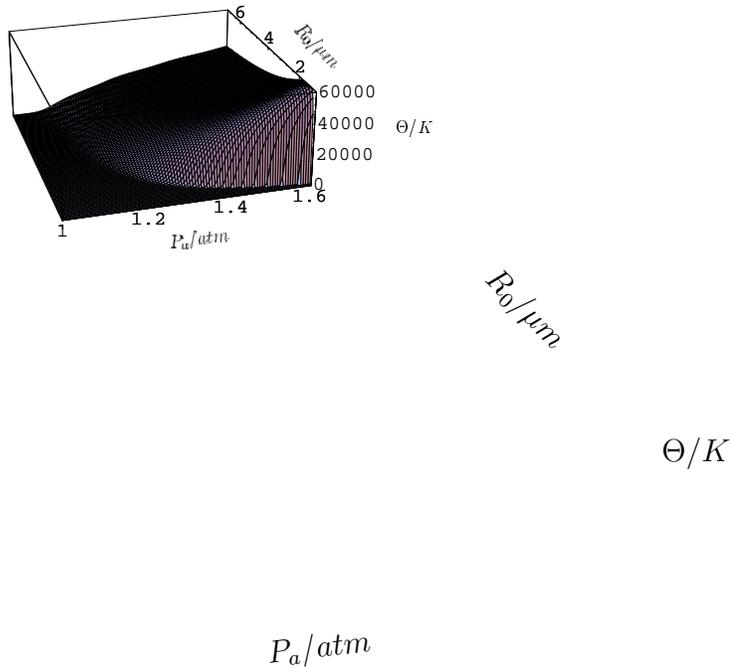,width=12cm,angle=0.}}
\end{picture}
\caption[Maximal temperature in the bubble]{
Maximal temperature in the bubble as a function of $R_0$ and $P_a$
within
our model calculation.
}
\label{5temp}
\end{figure}

What are the temperatures achieved in this approach? In figures 
\ref{rp_r02and5}d
and
\ref{5temp} we plot the {\it maximal} temperature within our model as a
function of $R_0$ and $P_a$.
In the parameter regime of interest for SL, values as high as
$20000-60000K$ are achieved. These values agree 
 order of magnitude wise with
the independent results of Bernstein et al. \cite{ber95,ber96} who
extracted the temperature from the spectral shape and obtained
30000 to 60000K, depending on the gas contents.

\section{Phase diagrams for air bubbles}
Now consider a bubble in water
containing a mixture
of a reactive gas (taken to be N$_2$) and an inert gas Ar.
The total number of moles of gas in the bubble is 
\be 
N_{tot} ={ 4\pi R_0^3 P_0 \over 3G\Theta_0  }= N_{N_2} + N_{Ar},
\label{5ntot}
\ee
where 
$\Theta_0 = 273$K is the normal temperature
and $G=8.3143J/(molK)$ is the gas constant.
The  argon ratio {\it in} the bubble is
\be \xi_b = {N_{Ar}\over N_{tot}}, \ee
that of nitrogen \be 1-\xi_b = {N_{N_2}\over N_{tot}}.
\ee 
If $c^{Ar}(r,t)$ and $c^{N_2}(r,t)$ are the concentration fields of
$Ar$ and $N_2$ in the liquid, respectively,
the rate of change of the
number of moles
of $N_2$ and $Ar$ in the bubble is given by
\begin{eqnarray}
\dot{N}_{Ar} &=& {4\pi R^2 D_{Ar} \partial_r c^{Ar}\vert_{r=R}\over
  \mu_{Ar}}
\label{ar}\\
\dot{N}_{N_2}&=&  {4\pi R^2 D_{N_2} \partial_r c^{N_2}\vert_{r=R}\over
\mu_{N_2}}
- A N_{N_2} 
 \exp{\left(- {\Theta^*\over \Theta}\right)}\label{n2}.
\end{eqnarray}
Here, $D_{Ar}$, $D_{N_2}$,
$\mu_{Ar}$ and $\mu_{N_2}$ are the respective
diffusion constants and molecular masses.
The concentration fields obey a  mass advection diffusion equation
\cite{fyr94,hil96},
whose boundary conditions are set by the external concentrations
$$c^\alpha (\infty , t) = c_\infty^\alpha
=
c_0^\alpha {
p_\infty^\alpha \over P_0}
$$ (Henry's law)
and by the partial gas
pressures $p^\alpha (t) $ in the bubble
$$c^\alpha (R(t),t)
=
c_0^\alpha
{p^\alpha (R(t))\over P_0},
$$
 $\alpha =$  Ar,  N$_2$.
The solubilities for nitrogen and argon are different,
$c_0^{Ar} = 0.061 kg/m^3$ and
$c_0^{N_2} = 0.020kg/m^3$ \cite{lid91}. The diffusion constants are
approximately the 
same \cite{lan69}, $D_{Ar} = D_{N_2}=
2\cdot 10^{-9} m^2/s$.
The second term in (\ref{n2}) represents the bubble's nitrogen
loss
by chemical reaction.
The reaction rate will depend on the temperature $\Theta (t)$ in the
bubble.
For simplicity,
we assume that the reactions follow an Arrhenius law, with
empirical parameters  appropriate for nitrogen dissociation
(ref.\ \cite{ber96}):
$A=6 \cdot 10^{19} (\Theta_0 / \Theta)^{2.5}
(\rho_0/\mu_{N_2})(R_0/R)^3cm^3/(mol \cdot s)$
giving the timescale
of the reaction;
$\Theta^* = 113000K$ \cite{ber96}
is the activation temperature and
 $\rho_0$
the equilibrium gas density.
This reaction law is rather crude, as it
neglects backward reactions as well as the kinetics of the 
expulsion of reaction products; however, 
it is sufficient for this demonstrative calculation.

We can straightforwardly extend the adiabatic approximation
of the
slow diffusional dynamics \cite{fyr94,loe95,hil96}
(i.e., separation of times scales)
to 
equations (\ref{ar}), (\ref{n2}). The only requirement is that the
involved chemical reactions are fast compared to diffusional processes
which definitely is the case. The result of the adiabatic
approximation is that
the change per cycle 
is given by
\begin{eqnarray}
{\Delta N_{Ar} \over T} &=& {4\pi  D_{Ar} c_0^{Ar} 
\over \mu_{Ar}  P_0 I } 
\left( 
p_\infty^{Ar} - \xi_b
\left< p \right>_{4}
\right)
\label{ndot_ar}
\\
{\Delta N_{N_2}\over T} &=& {4\pi  D_{N_2} c_0^{N_2}
\over \mu_{N_2} P_0 I } 
\left( 
p_\infty^{N_2} -  (1-\xi_b)
\left< p \right>_{4} 
\right) -
 {N_{N_2}} \left< A
\exp{\left(- {\Theta^* /  \Theta}\right)}\right>_{0} 
\label{ndot_n2}
\end{eqnarray}
with the 
 {\it weighted} time averages
\begineq
\left<f(t)\right>_{i} =
{\int_0^T f(t) R^i(t) dt \over
\int_0^T R^i(t) dt}.
\label{av}
\endeq
For further simplification in the numerical calculations to come
we in addition employ the saddle point approximation of
the integral $I$ defined by
\begineq
I= \int_0^\infty {dh' \over \la ( 3h' + R^3(t))^{4/3}\ra_{0}}
\label{int}
\endeq
as 
$I\approx 1/R_{max}$, 
as already done in \cite{loe95}.

The last term in eq.\ (\ref{ndot_n2}) will only contribute at the
collapse when $\Theta(t)$ is large. The chemical reaction rate will
depend on the detailed space and time dependence of the temperature in
the bubble which we model
by the polytropic law (\ref{2temp})
-- as explained in the previous section.

\begin{figure}[htb]
\setlength{\unitlength}{1cm}
\begin{center}
\begin{picture}(11,9)(0.2,-0.5)
\put(-1.5,-0.3){\psfig{figure=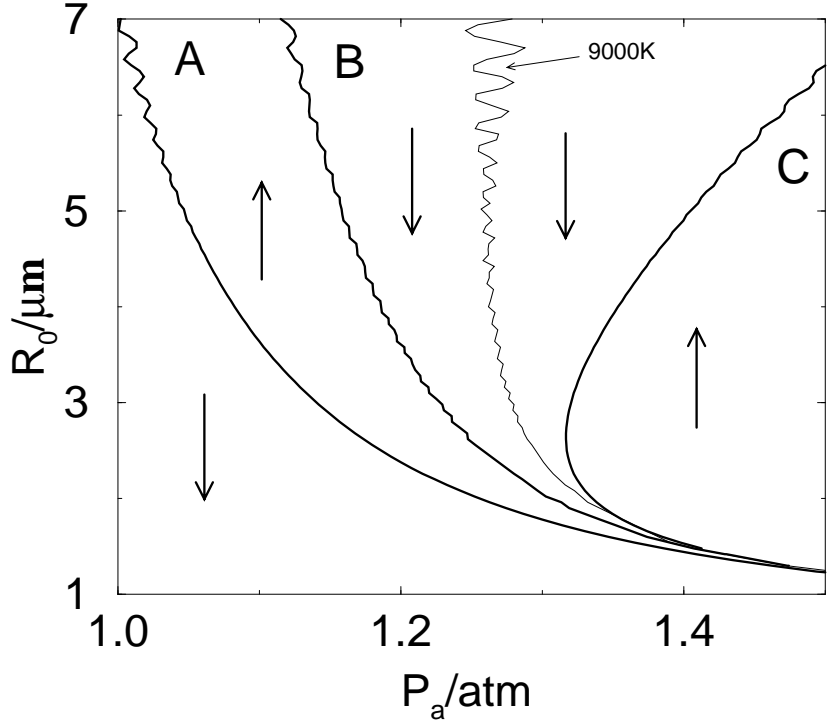,width=13.5cm,angle=-90}}
\end{picture}
\end{center}
\caption[$R_0$- $P_a$ phase diagram for air at $p_\infty/P_0=0.20$  ]{
Phase diagram for air at $p_\infty/P_0=0.20$
in the $R_0$ - $P_a$ space.
The arrows
denote whether the ambient radius grows or shrinks at this parameter
value.  Curve A denotes the  
equilibrium for an air bubble, on curve C the bubble only contains 
argon.   The intermediate curve B
necessarily exists because of the topology of the diagram, and represents
an additional stable equilibrium.
The thin line shows when the nitrogen dissociation threshold $9000K$ is
reached.
}
\label{ourholt}
\end{figure}

With these approximations 
equilibrium points 
$\Delta N_{Ar} =\Delta N_{N_2} = 0$ 
in the two dimensional
space ($N_{Ar}, N_{N_2}$), or equivalently in the space ($\xi_b ,
R_0$), 
can easily be calculated from the RP dynamics (\ref{rp}) via equations 
(\ref{ndot_ar}), (\ref{ndot_n2}).
As in ref.\ \cite{hil96},
only averages of type (\ref{av}) have to be determined for each
parameter pair $(R_0,P_a)$; therefore, phase diagrams can
be calculated solely from the Rayleigh-Plesset equation (\ref{rp}). 
The equilibrium radii $R_0^*$ 
in the $R_0-P_a$ plane for air ($\xi_l = 0.01$) at $p_\infty / P_0 =0.20$
are shown in figure \ref{ourholt}. 
For small forcing the
temperatures
are not high enough to initiate chemical reactions, so
that the equilibrium curve corresponds to the prediction
of ref.\ \cite{hil96} for this gas concentration.
This equilibrium is unstable: The bubble either shrinks or
grows by rectified diffusion. As pointed out above the  
growing bubble
eventually
runs into a shape instability where
microbubbles pinch off
and make the bubble dance because of the recoil
\cite{hil96}. 
In the opposite limit of high forcing (curve C),
the reactions
burn off all the N$_2$, so that
the bubble contains pure argon; this equilibrium corresponds to the
(stable) equilibrium
at the argon partial pressure $p_\infty^{Ar}/P_0 = 0.01 p_\infty/P_0=0.002$.

Figure \ref{ourholt} displays a regime of shrinking bubbles at high forcing
pressures (left of curve C) and an adjacent region of growing bubbles
(right of curve A).
This necessitates the existence of an additional
equilibrium at intermediate forcing pressures,
curve B in figure \ref{ourholt}, for which growth by rectified diffusion
and mass loss by reactions balance.  This 
additional equilibrium occurs
close to the point of nitrogen dissociation,
and turns out to be stable; the argon fraction 
$\xi_b^*$ for this equilibrium is
slightly larger than the fraction $\xi_l$ in the liquid (for
not too strong forcing).

\begin{figure}[htb]
\setlength{\unitlength}{1cm}
\begin{center}
\begin{picture}(11,9.5)(0.2,-0.5)
\put(-2.2,-1.0){\psfig{figure=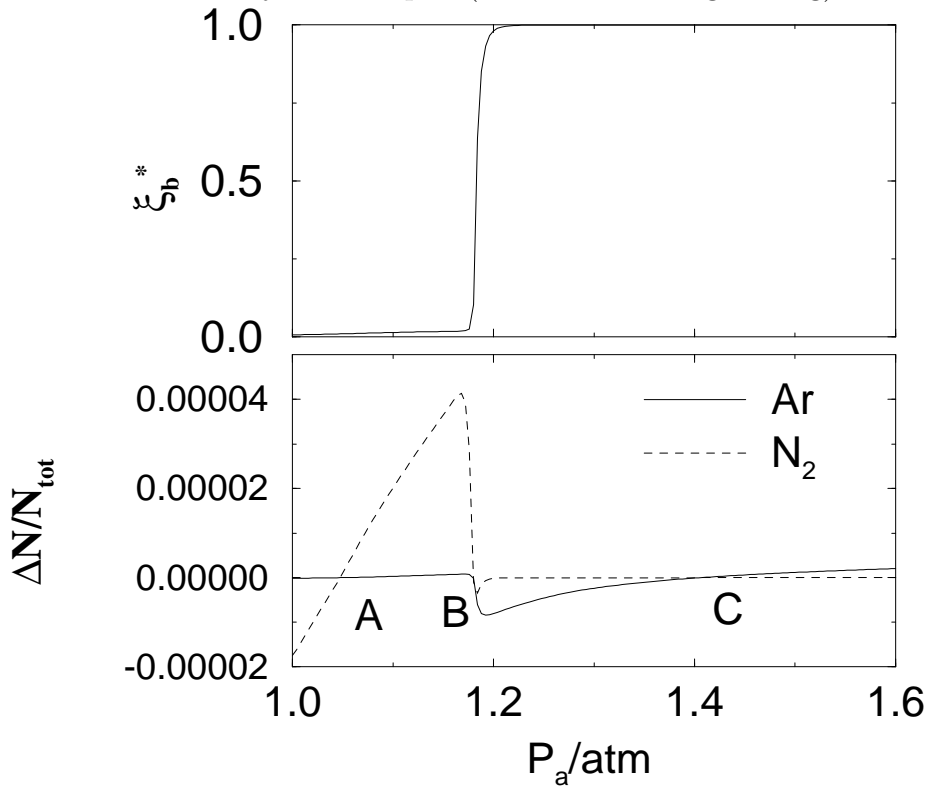,width=13.6cm,angle=-90}}
\end{picture}
\end{center}
\caption[Balance of argon and nitrogen in the bubble]{
The top curve shows the equilibrium argon ratio $\xi_b^*$
in a $R_0=5\mu m$ bubble as a function of the forcing pressure $P_a$.
The bottom curve presents the corresponding
argon and nitrogen loss/gain per cycle,
normalized to the total amount of moles. 
}
\label{chem_bal}
\end{figure}

In figure \ref{chem_bal} we present the net loss/gain per cycle for
both argon and nitrogen for a $R_0=5\mu m$ bubble. At small 
$P_a < 1.15$atm the second term in equation (\ref{ndot_n2}) is not important as
the gas does not become hot enough. The bubble diffusively
shrinks (for $P_a<1.045$atm) or grows because of rectified diffusion;
the argon ratio is constant at $\xi_b^* = \xi_l = 0.01$. In between
the shrinking and the growing regime is the unstable equilibrium point A.
Around  
$P_a = 1.17$atm
the heating is sufficient to lead to some nitrogen dissociation and
at $P_a= 1.18$atm the nitrogen loss through dissociation balances
its gain through rectified diffusion. At that forcing pressure also argon is in
equilibrium as the argon ratio has increased up to $\xi_b^* \approx 0.1$
so that $p_\infty^{Ar} = \xi_b^* \left< p\right>_{4}$: We have reached the
stable equilibrium point B. At larger forcing pressure
$P_a$ the bubble is in the dissolution
island. For increasing $P_a$ the nitrogen loss becomes
less because less and
less nitrogen is left in the bubble. The argon loss, on the other hand,
 diminishes because we approach the stable diffusive equilibrium point C
 at $P_a\approx 1.4$atm. Beyond this forcing pressure, the bubble grows again
 by rectified argon diffusion; essentially no nitrogen is left in the bubble
$ (1-\xi_b^* \approx 6\cdot 10^{-5}$ at $P_a = 1.40$atm).
Note that because of the separation of time scales the
values on $\xi_b^*$ refer to an {\it average} value over the whole
cycle.
When the bubble is close to its maximum, the nitrogen concentration will
be slightly larger, right at the collapse the argon concentration will
be slightly larger.

\begin{figure}[htb]
\setlength{\unitlength}{1.0cm}
\begin{center}
\begin{picture}(11,6.5)(0.3,2.2)
\put(-2.,-8.3){\psfig{figure=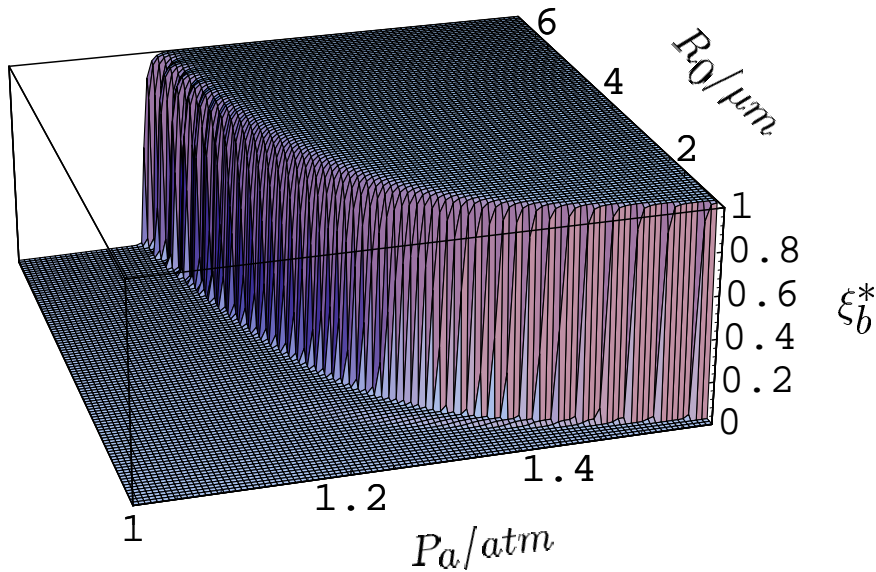,width=16cm}}
\end{picture}
\end{center}
\caption[Fraction of argon in the bubble]{
The fraction
$\xi_b^*$ of argon in the bubble
as a function of $P_a$ and $R_0$.
}
\label{xi_fig}
\end{figure}

How robust is this picture, i.e., how does it depend on details of 
temperature dynamics and the chemical reaction mechanism which are
both poorly understood? The large $R_0$ parts of branches
 A and C do not depend at all on these
details. The reason becomes clear from
figure \ref{xi_fig} in which we show the composition equilibrium $\xi_b^*$
as a function of $R_0$ and $P_a$. 
It is given by 
\be {\Delta N_{Ar}(\xi_b^*)
\over \Delta N_{N_2} (\xi_b^*)} = {\xi_b^* \over 1-\xi_b^* }.
\label{5equil}
\ee
Plugging (\ref{ndot_ar}) and (\ref{ndot_n2}) 
into (\ref{5equil}) gives the quadratic equation
for the equilibrium argon ratio $\xi_b^*$
\be
(\alpha + \epsilon ) \xi_b^{*2} + (\beta - \epsilon )
 \xi^*_b + \delta = 0 \label{5quadeq}
\ee 
with the coefficients
\begin{eqnarray}
\alpha &=&  
{4\pi \over  I} 
{\langle p \rangle_{4} \over P_0}
\left(  
{D_{N_2} c_0^{N_2} \over \mu_{N_2}} -
{D_{Ar} c_0^{Ar} \over \mu_{Ar}} \right) ,
\label{alpha}
\\
\beta &=& {4\pi \over  IP_0}  \left(
{D_{N_2} c_0^{N_2} \over \mu_{N_2}} \left( 
{p_\infty^{N_2} } - {\langle p \rangle_{4}
  }\right) +
{D_{Ar} c_0^{Ar} \over \mu_{Ar}} \left( 
{p_\infty^{Ar} } + {\langle p \rangle_{4}
  }\right) \right) ,
 \label{beta}
\\
\epsilon &=& {4\pi P_0 R_0^3 \over 3 G \Theta_0 } \left<A \exp \left(
-{\Theta^*\over \Theta } \right) \right>_{0},
\label{epsilon}
\\
 \delta &=& -{4\pi D_{Ar} c_\infty^{Ar} \over I \mu_{Ar}}.
 \label{delta}
 \end{eqnarray}
From the two solutions only the one with $0 \le \xi_b^* \le 1$ has
physical meaning.
The interpretation of figure \ref{xi_fig} is straightforward:
 Weakly forced bubbles
have $\xi_b^* \approx \xi_l$, thus $p_\infty/P_0 = 0.20$ 
is relevant for stability.
Strongly forced bubbles have $\xi_b^*
\approx 1 \gg \xi_l$, thus $p_\infty^{Ar} /P_0= 0.002$ is the relevant
quantity.
The transition between these regimes is abrupt, and occurs when the bubble
temperature surpasses the dissociation temperature ($\approx 9000K$ for N$_2$).
Where exactly this happens depends on the model of the temperature and
thus so does the equilibrium curve B in figure \ref{ourholt}.

However, the {\it existence} of this additional stable equilibrium is
independent of the model details. It simply follows from topological
reasons: To the right of the curve A the bubbles are growing, to the
left of curve C
they are shrinking, so in between there must be an 
equilibrium. Indeed, as mentioned already above
and shown in figure \ref{holt}, Holt and Gaitan's \cite{hol96} 
recent detailed measurements of $R_0-P_a$ phase diagrams found such a
classically unexpected equilibrium (for the same air pressure
$p_\infty / P_0 = 0.20$ as chosen here) and the connected large $P_a$
``dissolution island'' between curves B and C.

\begin{figure}[htb]
\setlength{\unitlength}{1cm}
\begin{center}
\begin{picture}(11,9.5)(0.2,-0.5)
\put(-1.5,-0.5){\psfig{figure=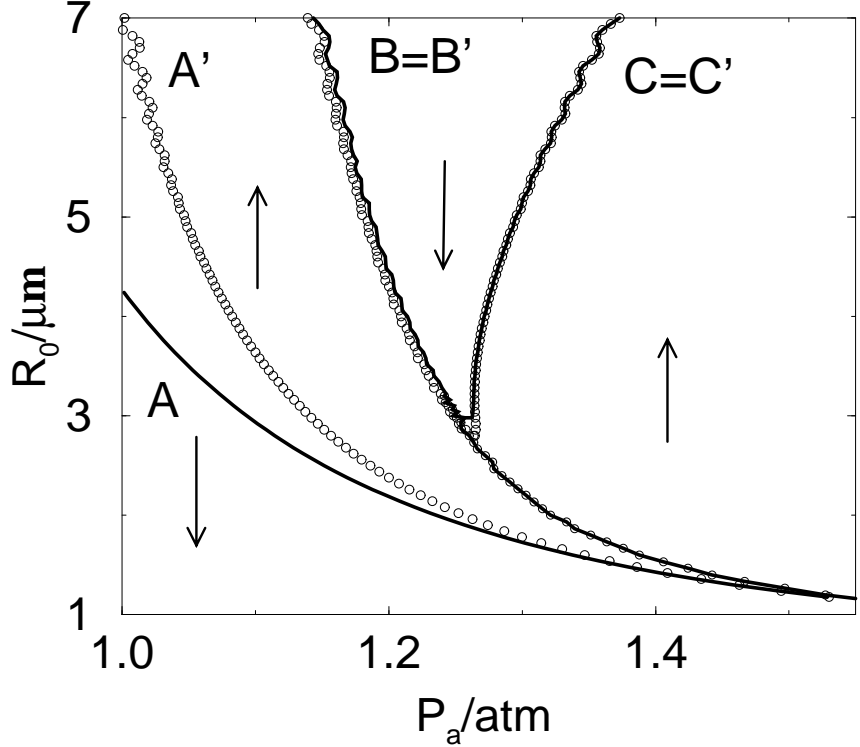,width=14cm,angle=-90}}
\end{picture}
\end{center}
\caption[$R_0$- $P_a$ phase diagram for air at $p_\infty/P_0=0.50$  ]{
The solid lines (curves A, B, C)
show the same phase diagram for air
as in figure \ref{ourholt}, but now for
$p_\infty/P_0=0.50$, i.e.,
$p_\infty^{Ar}/P_0=0.005$. For comparison, the open symbols
(curves A$'$, B$'$, C$'$) show
the phase diagram for a nitrogen-argon mixture with
an argon ratio of $\xi_l = 0.025$ at 
$p_\infty/P_0=0.20$, which beyond the nitrogen dissociation
threshold results in the same 
partial argon pressure $p_\infty^{Ar}/P_0=0.005$.
Indeed, for large $P_a$ the phase diagrams agree.
Only curves A and A$'$ below the nitrogen dissociation threshold
differ.
}
\label{ourholt_xi01p50}
\end{figure}

Holt and Gaitan \cite{hol96} repeated 
the experimental phase space measurements for larger air pressures
$p_\infty/P_0 = 0.40$
and
$p_\infty/P_0 = 0.50$.
The { theoretical} phase diagram for 
$p_\infty/P_0 = 0.50$ and $\xi_l = 0.01$ (air)
is shown in figure \ref{ourholt_xi01p50}.
The island of dissolution between curves B and C is now smaller, as the partial
argon pressure is $p_\infty^{Ar}/P_0 = 0.005$ and curve C is thus shifted to
the left, whereas the nitrogen dissociation curve
is hardly affected
and consequently neither are curves  A and B. In
experiment we see the same {\it qualitative} behavior, but even stronger
pronounced, see figure 1a of ref.\ \cite{hol96}. We expect the poor modeling
of both the temperature dependence and the chemistry as the origin of the 
quantitative disagreement of the theoretical and experimental phase diagrams.

For comparison we also calculated the phase diagram for $p_\infty/P_0 = 0.20$
and $\xi_l = 0.025$ and plotted it into the same figure \ref{ourholt_xi01p50}.
For this situation the argon percentage $p_\infty^{Ar}/P_0 = \xi_l p_\infty/P_0
= 0.005$ is the same as for $p_\infty/P_0 = 0.50$, $\xi_l=0.01$. Indeed, the
phase diagram demonstrates very convincingly
that for large forcing pressure it
is only the
partial argon pressure which is the relevant parameter for diffusive
stability: The equilibrium curves B and C for the two cases agree very well.

\begin{figure}[htb]
\setlength{\unitlength}{1cm}
\begin{center}
\begin{picture}(11,9.5)(0.2,-0.5)
\put(-1.7,-0.3){\psfig{figure=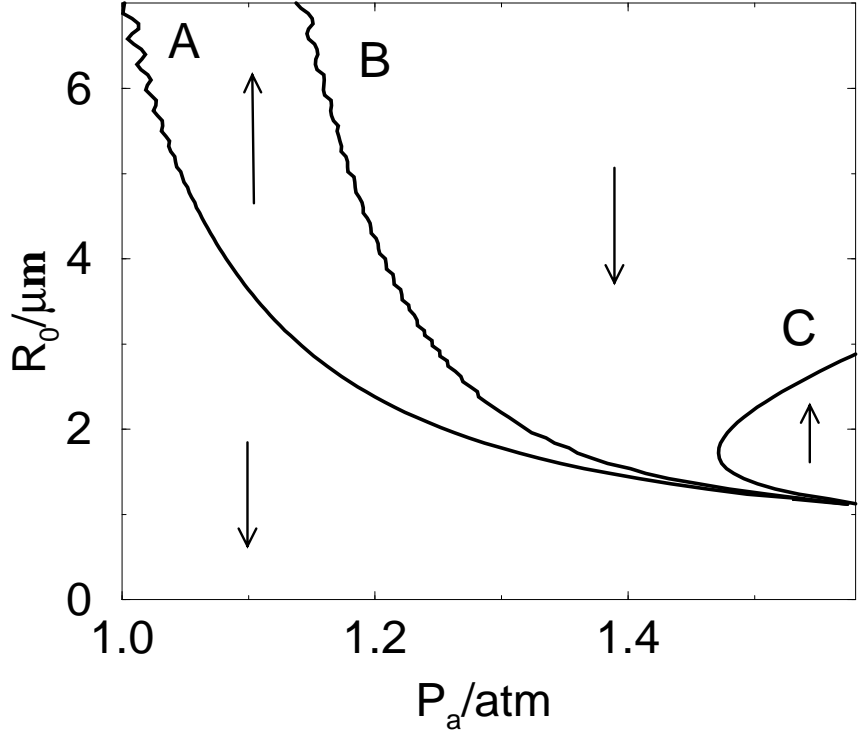,width=14cm,angle=-90}}
\end{picture}
\end{center}
\caption[$R_0$- $P_a$ phase diagram for $p_\infty/P_0=0.20$, 
$\xi_l=0.001$]{
$R_0$ - $P_a$
phase diagram for nitrogen with a small amount
$\xi_l= 0.001$ of argon dissolved.
The relative pressure overhead is
$p_\infty/P_0=0.20$. These parameters correspond to the experimental
situation 
in figure 11 of L\"ofstedt et al.\ \cite{loe95}, see figure
\ref{fig_gap} of the present work.
}
\label{ourholt_xi001p20}
\end{figure}

\section{Phase diagrams for further nitrogen-argon mixtures}
Our theoretical method can be applied to many other experimental situations.
As a further
example we calculate the corresponding phase diagram
figure \ref{ourholt_xi001p20} to
the experimental situation in figure 11 of \cite{loe95},
see figure \ref{fig_gap} here, 
where $p_\infty/P_0=0.20$ and $\xi_l=0.001$.
The total gas concentration
$p_\infty/P_0$ has the same value as in figure \ref{ourholt}. Therefore,
the curves A and B
are also essentially as in that figure, because they are determined by the
nitrogen concentration. Curve C which is beyond the
nitrogen dissociation threshold, however, is further to the right,
reflecting the lower argon concentration $p_\infty^{Ar}/P_0= 0.0002$.

What happens if $P_a$ is slowly increased? For small $P_a=1$atm
the bubble
will be repelled from the unstable equilibrium A. It either dissolves or
grows by rectified diffusion and runs into the shape instability where
it pinches off a microbubble and the whole process starts over, as described
in detail in ref.\ \cite{hil96}. For $P_a\approx 1.15$atm 
the bubble does not run into the shape instability any more but in the
attractive diffusive equilibrium curve B where it stays. On further increase
of $P_a$ it follows that curve and shrinks drastically.
Exactly this
sequence of
events has experimentally been observed,
see figure 11 of L\"ofstedt et al.\ \cite{loe95}.

For larger $P_a$  on first sight there seems to be a contradiction to the
experimental observation figure
11 of \cite{loe95}. Theoretically, the bubble should
always stay on the {\it stable} equilibrium curve B. Experimentally, however,
between $P_a=1.3$atm and $1.4$atm no bubble seems to exist and at $P_a=1.4$atm
the ambient radius is much larger, similar to that of curve C rather than that
of curve B.

This variance may be resolved when we consider the
unavoidable {\it jitter} in the forcing
pressure amplitude $P_a$. At $P_a=1.35$atm, the equilibrium curves 
A and B are so close that a small jitter $\Delta P_a$
(to smaller $P_a$) of less than
$0.05$atm (which is less than the precision to which $P_a$ can be determined;
also the translational movements of the bubble
and its distortion on the pressure field  \cite{mat97} could
contribute to a jitter.)
suffices to make the bubble jump to the unstable domain below
curve A where it dissolves. Thus, though the equilibrium B in that regime is
stable from a mathematical point of view, from a physical point of view it is
not and the bubble can 
dissolve.
Support for this claim could come from an experiment in which the jitter
is artificially increased: the gap in figure \ref{fig_gap}
where no bubble exists should widen.

With the help of the phase diagram
figure \ref{ourholt_xi001p20} we now better
understand the above mentioned hysteresis.
The stable part of branch C is reached by a fast increase of $P_a$. The bubble
is then boosted into the growing regime { below}
 the stable branch of curve C. Subsequently, this stable branch C
is reached through rectified diffusion. On the other hand, if $P_a$ is
decreased from $P_a=1.4$atm, the bubble will  first
follow the stable equilibrium curve C and
shrink. At the bifurcation point it falls off the equilibrium curve
and shrinks towards the stable
curve B.
Because of the jitter in $P_a$ curve B may not be able to trap the
bubble as explained above and the bubble dissolves. 
Even a fast decrease of $P_a$
towards much smaller values $\approx 1.1$atm 
does not save the bubble from dissolution 
as it only contains argon and the nitrogen needed for a stable equilibrium
has been burned off in the first part of the cycle during which
$P_a$ was increased and recollecting nitrogen only happens on the slow
diffusive timescale.

\begin{figure}[p]
\setlength{\unitlength}{1.0cm}
\begin{center}
\begin{picture}(11,18.8)
\put(-1.5,10.1){\psfig{figure=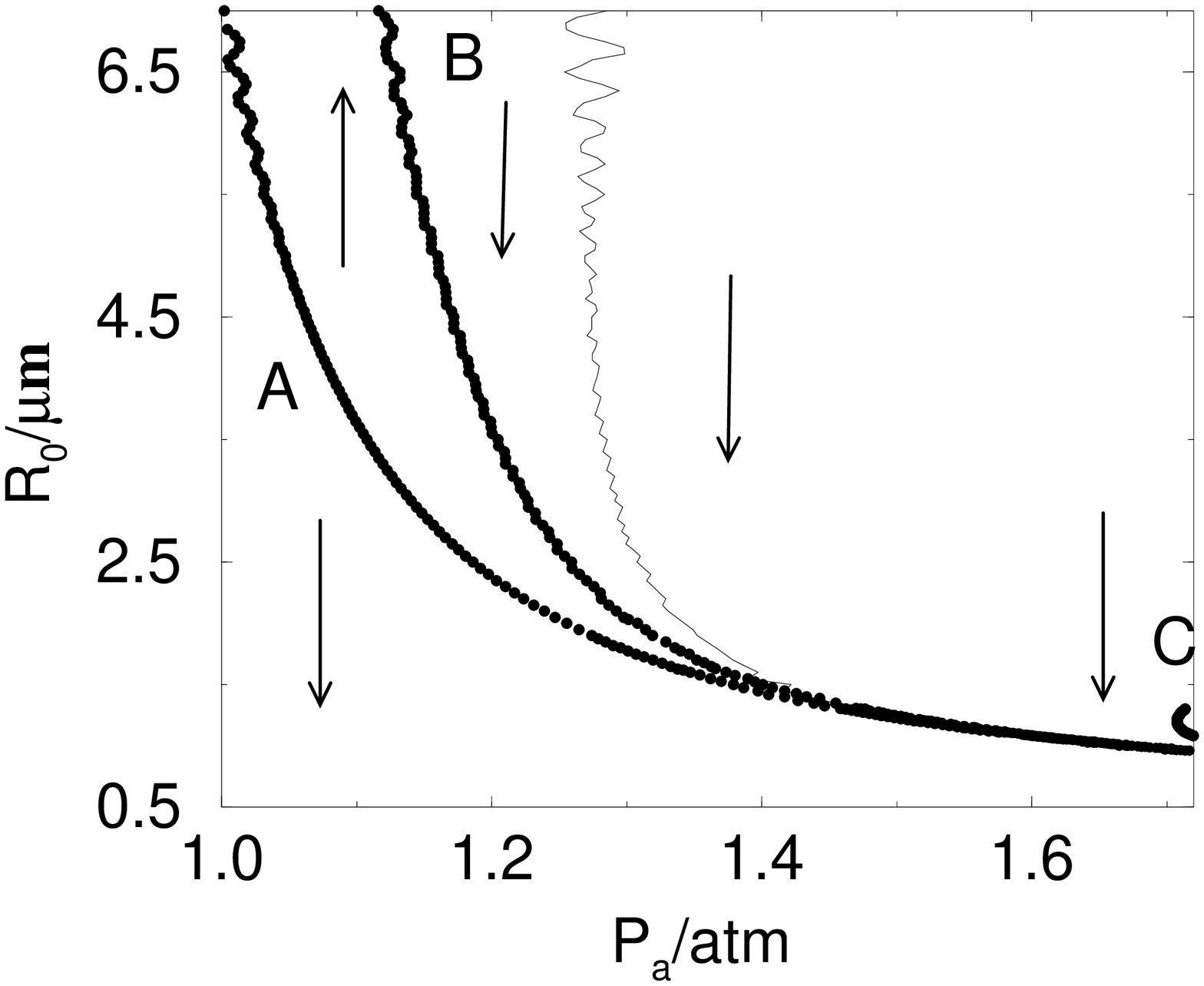,width=12.5cm,angle=-90}}
\put(-1.5,0.5){\psfig{figure=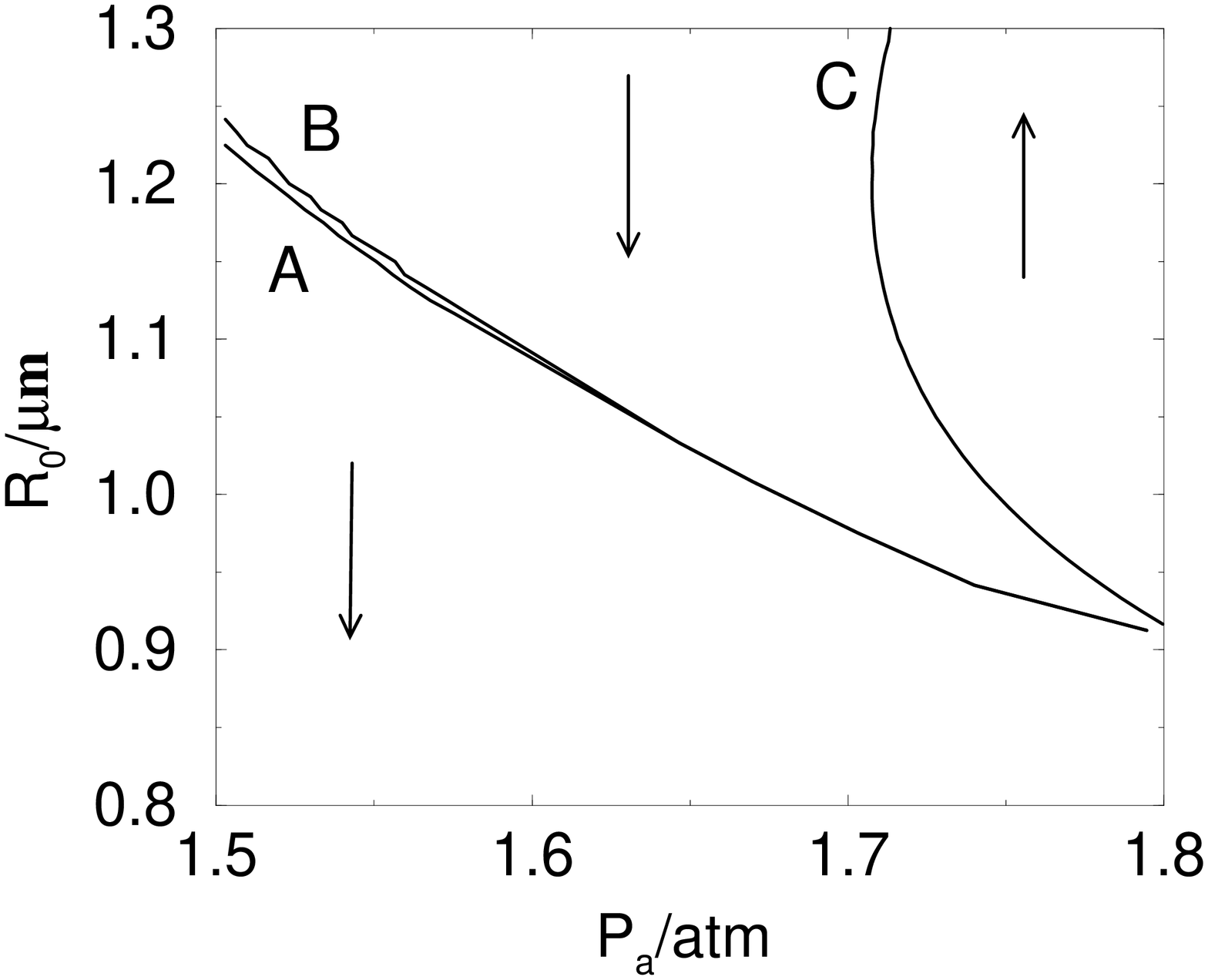,width=12.5cm,angle=-90}}
\end{picture}
\end{center}
\caption[$R_0$- $P_a$ phase diagram for $p_\infty/P_0=0.20$, 
$\xi_l=0.0001$]{
Phase diagram in $R_0$-$P_a$ space for $p_\infty/P_0=0.20$ and
a tiny argon concentration of $\xi_l=0.0001$. Only for very high
forcing pressure diffusively stable bubbles are possible
(but are likely to be unstable towards the surface shape instability).
 The bottom 
figure shows an enlargement of the lower right part of the top figure. 
}
\label{ourholt_xi0001p20}
\end{figure}

For even lower argon concentration as e.g.\ in figure \ref{ourholt_xi0001p20}
where we chose $\xi_l=0.0001$ and $p_\infty/P_0=0.20$ the situation is 
similar. The only difference is that curve C moves even
further to the
right and may not be reached in experiment because of the (short
timescale) shape instabilities discussed in refs.\ \cite{bre95,hil96}.

Let us discuss the situation in the limiting case
$\xi_l=0$, i.e., pure nitrogen. As explained above for $\xi_l=0.001$,
on increasing $P_a$ the bubble should first ``dance'' because of
growth by rectified diffusion and pinching off of microbubbles and then
it should be caught by the {\it stable} (but not necessarily sonoluminescing)
equilibrium curve B.
At this equilibrium bubble growth by rectified diffusion and
nitrogen loss by dissociation again just balance.
For large 
$P_a \sim 1.3$atm the bubble  should dissolve by the same mechanism discussed
above: Curves A and B are so close that a jitter in $P_a$ leads to a jump
from the stable equilibrium curve B to the dissolution domain below
the unstable curve A.
We suggest to experimentally look for the stable
equilibrium curve B which is suggested by the nitrogen dissociation
hypothesis.

What experiments on pure nitrogen bubbles have been done up to now?
To our knowledge
only one experiment has been reported (Hiller et al.\  \cite{hil94}).
In that experiment
the nitrogen was $99.7\%$ pure. Hiller et al.\
\cite{hil94} managed to measure extremely weakly
sonoluminescing unstable bubbles which were hard to keep alive.
This supports the presented theory that in the strong forcing regime the high 
temperature destroys the nitrogen and thus the bubbles.
However, the light intensity figure 5 of ref.\ \cite{hil94} shows an
oscillating  pattern (on a timescale of seconds) 
which we do not understand.

\section{Inert gas mixtures}
\begin{figure}[htb]
\setlength{\unitlength}{1.0cm}
\begin{center}
\begin{picture}(11,6.5)(0.3,2.2)
\put(10.3,5.0){$\xi_b^*$}
   \put(8.4,7.9){
      \begin{rotate}{45}
             $R_0/\mu m$
        \end{rotate}}
    \put(5.6,2.5){
        \begin{rotate}{-7}
             $P_a/atm$
        \end{rotate}}
\put(-1.5,-3.3)
{\psfig{figure=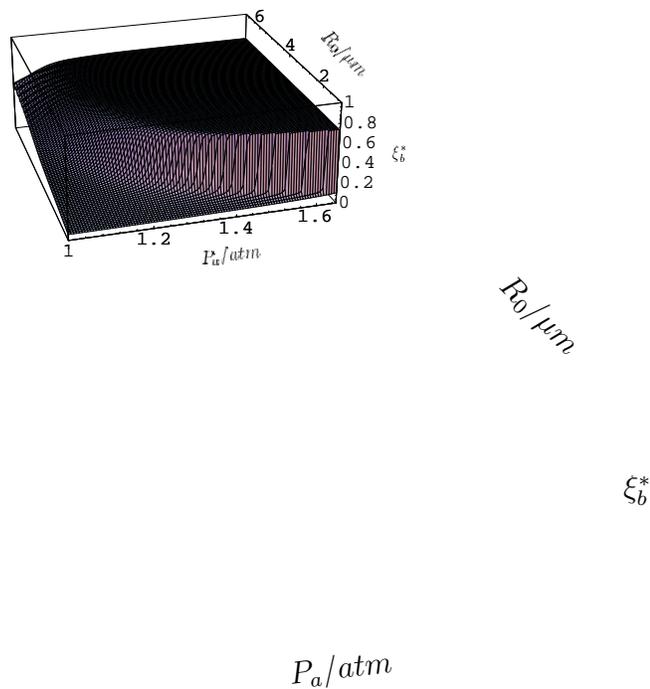,width=12cm,angle=0}}
\end{picture}
\end{center}
\caption[Argon accumulation in the bubble for an argon-helium mixture]{
Argon equilibrium
ratio $\xi_b^{*}$ in a bubble as a function of ambient radius $R_0$
and forcing pressure $P_a$. The external gas is a 1:1 mixture
of argon and helium at an overhead pressure of $p_\infty/P_0 = 0.20$.
Only the difference in the diffusion constants and in the
equilibrium concentrations between argon and helium leads to the
argon accumulation in the bubble for large forcing.
}
\label{xi_he_ar_xi50p20}
\end{figure}

It is interesting to note that the chemical instability of one
component of the gas mixture is not a necessary requirement for
the accumulation of one gas species in the bubble. A similar accumulation
can be achieved for inert gas mixtures
if the diffusion constants and the equilibrium
concentrations of the two gases in the mixtures are different.
As an example we consider a 1:1 mixture
of helium and argon at $p_\infty/P_0
=0.20$, i.e.,
$\xi_l=0.50$ for the argon ratio.
The diffusion constants are
$D_{Ar}= 2\cdot 10^{-9} m^2/s$ and 
$D_{He}= 5.8\cdot 10^{-9} m^2/s$ \cite{lan69} and 
the equilibrium concentrations are
$c_0^{Ar}=61 g/m^3$ and 
$c_0^{He}=0.78 g/m^3$ \cite{lid91}.
The molecular masses are $\mu_{Ar} = 40 g/mol$ and $\mu_{He}=4g/mol$.
The resulting equilibrium concentration $\xi_b^*$
in the bubble is shown in figure \ref{xi_he_ar_xi50p20}. For small
forcing helium accumulates in the bubble and 
for large forcing argon accumulates.
The borderline between these two regimes
is the unstable diffusive equilibrium curve calculated in 
\cite{hil96} and denoted ``curve A'' within the plots shown here. 
As no reactions occur, this is the only equilibrium: Below that curve
bubbles diffusively shrink, above it they grow by rectified diffusion.
{}From this behavior the helium or argon accumulation can
immediately be qualitatively understood: The diffusive change
of mass is proportional
to the product of the material constants
$Dc_0/\mu$, see eq.\ (\ref{ndot_ar}).
Now 
$D_{Ar}c_0^{Ar}/\mu_{Ar} = 3.1\cdot 10^{-9} mol s^{-1}m^{-1}$
and 
$D_{He}c_0^{He}/\mu_{He} = 1.1\cdot 10^{-9} mol s^{-1}m^{-1}$.
Thus in the shrinking regime argon can escape faster and helium accumulates.
Vice versa, in the growing regime the rectified diffusion of argon 
into the bubble wins and it accumulates.
In the process of helium accumulation the helium
equilibrium concentration
$1-\xi_b^*$ may never be achieved as the bubble dissolves earlier.
This problem does not occur for the process of argon accumulation,
as the diffusively growing bubble
runs into the shape instability and can
survive after microbubble pinchoff \cite{hil96}.

The maximal argon concentration achieved is
only $\xi_{b,max}^*\approx 0.73$, less than the result
$\xi_{b,max}^* =1$ of the previous section
where the chemical instability of
nitrogen caused its complete extinction.
Also the transition in fig.\ \ref{xi_he_ar_xi50p20} is less abrupt
than in above case fig.\ \ref{xi_fig} 
for which chemical reactions occurred.

The maximal argon concentration can easily be calculated 
analytically from equations (\ref{5quadeq}) -- (\ref{delta})
with the index N$_2$ replaced by the index He. 
Obviously, $\epsilon = 0$ as no reactions take place.
{}From figure 7 of \cite{hil96}
we realize that for large forcing $\left< p\right>_{4}$
is small compared to
$p_\infty^{Ar}=
p_\infty^{He}=0.1$. With this large forcing pressure
approximation we obtain
\be
\xi_{b,max}^* = {
{D_{Ar} c_\infty^{Ar}\over \mu_{Ar}}
\over
{D_{Ar} c_\infty^{Ar}\over \mu_{Ar}} +
{D_{He} c_\infty^{He}\over \mu_{He}}
}
\approx 0.73
\label{hear}
\ee
in agreement with the numerical result figure \ref{xi_he_ar_xi50p20}.

\section{Conclusion, predictions, and suggested experiments}

We conclude this paper
with a summary of the nitrogen dissociation theory \cite{bre96b,loh97}
which we elaborated here in detail.
It accounts for the dependence of SBSL on the percentage of inert gas
in the bubble.
It is based on
a combination of principles from sonochemistry and hydrodynamic
stability.
The main result is
that
strongly forced air bubbles eventually consist of pure argon because
at the high temperatures achieved in the bubble
the  nitrogen and oxygen molecules dissociate and  react
to water soluble gases.
Consequently, it is the partial pressure 
$p_\infty^{Ar}$ of argon
(or of any other inert gas)
which is relevant for
stability \cite{loh97}.
The dynamics of the bubble is described by the Rayleigh-Plesset equation;
chemical reactions are assumed to obey an Arrhenius type law; back
reactions are neglected.
The bubble pressure is approximated by a van der Waals law.
The slow diffusive processes can be treated in an adiabatic approximation.
The theory contains no adjustable fit parameter.

Finally, we make 
detailed suggestions of experiments
to further test the nitrogen dissociation hypothesis.

\begin{itemize}
\item
{\it Phase diagrams in the $p_\infty/P_0$ vs $P_a/P_0$ parameter space:}
The most basic prediction is that in the large forcing regime the partial
pressure $p_\infty^{Ar}/P_0$ is the relevant parameter for obtaining stable
SL (or the partial pressure of any other inert gas). We suggest to map
out the phase diagram figure \ref{phase_dia_withdata} for 
argon-nitrogen mixtures at various  pressure overheads $p_\infty/P_0$ and
various argon ratios $\xi_l$: (i) Only the partial argon pressure
$p_\infty^{Ar}/P_0 = \xi_l p_\infty/P_0$
should matter and (ii)
the measured phase diagram should look like the theoretical
one figure \ref{phase_dia_withdata}.
\item
{\it Stable SL without degassing:}
In particular, stable SL will be possible without degassing, i.e., at
$p_\infty/P_0 = 1$, if the argon ratio is properly adjusted. For $P_a = 1.3$atm
it must be $\xi_l\approx 0.0033$. To test this prediction, extremely
pure water has to be used to avoid spontaneous cavitation.
An alternate way to achieve stable SL without degassing is to increase
the ambient pressure correspondingly.
\item
{\it Oxygen-argon mixtures:} Oxygen has a dissociation temperature
which is only
slightly lower than that of nitrogen. Therefore, oxygen-argon mixtures should
behave very similar
to nitrogen-argon mixtures. In particular, at large forcing only argon is left
in the bubble, so even the light intensities and spectra should be very
similar.
This prediction has recently been confirmed in experiment
\cite{cho96b}.
\item
{\it Other inert gas doped molecular gases:}
Apart from CO, nitrogen is the most stable molecular gas. Other molecular gases
dissociate earlier. Therefore, mixtures of them with inert gases will show a
slight shift of the stable equilibrium curve B towards smaller forcing pressure.
As the temperature increase with forcing pressure is very sudden (see
figure \ref{5temp}), this shift will, however, be very small and may be 
undetectable.
\item
{\it Stable nitrogen bubbles:}
Another very basic prediction of the nitrogen dissociation theory is the
existence of the equilibrium curve B in the phase diagrams of type
\ref{ourholt}. For air Holt and Gaitan have observed it \cite{hol96}.
We predict that it exists for any reactive gas.
In particular, we expect
that pure nitrogen
bubbles can be stable.
\item
{\it Island of bubble dissolution for large $P_a$:}
Between the equilibrium curves B and C, bubbles dissolve. The size of
this large $P_a$ island of dissolution should only depend on the partial argon
pressure $p_\infty^{Ar}/P_0 = \xi_l p_\infty/P_0$, hardly on the argon
ratio $\xi_l$ or the pressure overhead $p_\infty/P_0$ individually. For larger
$p_\infty^{Ar}/P_0$ 
the size of the dissolution regime should shrink, for smaller 
$p_\infty^{Ar}/P_0$ 
it should grow.
\item
{\it Radon doped SL bubbles:}
By using radon-nitrogen gas mixtures it may be possible to detect the
inert gas accumulation in the bubble because 
of the radioactivity of radon. 
\item
{\it Inert gas accumulation:}
Even for a mixture of nonreactive gases as He and Ar we in general have
$\xi_b \ne \xi_l$.
In the large forcing pressure regime
argon accumulates
for this example. 
\item
{\it pH measurements for air bubbles:}
A very appealing test is 
to measure the concentration
of the reaction products
of the dissociated gases
as a function of time, as already done in
MBSL \cite{har87}.
For nitrogen-argon mixtures 
nitrous acid production would
lead to a decrease in pH as already predicted in ref.\ \cite{loh97}. 
For an  estimate of the production rate we assume
that all the nitrogen that diffuses into the
bubble during the bubble expansion is burned off
at the collapse.
This amount is estimated in ref.\ \cite{loe95} as 
\be\Delta N_{N_2} =
{2\pi D_{N_2} c_\infty^{N_2} R_{max} T \over\mu_{N_2}}
\label{5ritva}
\ee
 per cycle. With typical
values of $R_{max} = 10 R_0$ for the maximal radius, $R_0 = 5\mu m$,
$D_{N_2}= 2\cdot 10^{-9} m^2/s$,
$c_\infty^{N_2} \approx 0.20 c_0^{N_2}$,
$c_0^{N_2}= 0.02 kg/m^3$, and $T=38\mu s$ one obtains 
$\Delta N_{N_2} \approx 3\cdot 10^{-18}$ mol per cycle or 
$\sim 3\cdot 10^{-10}$ mol of N$_2$ per hour converted to reaction
products. The consequence is
a small but detectable pH decrease: In a $100ml$
 flask with pure water (pH=7) there are initially $10^{-8}$ mol H$^+$
 ions. Assume that all ejected N atoms eventually form either HNO$_2$
 or HNO$_3$, then $6\cdot 10^{-10}$ mol H$^+$ ions are produced. This
 means a concentration increase to $(10^{-7} + 6\cdot 10^{-9}) mol/l$
 or a pH value of $6.975$ after an hour.
\item
{\it pH measurements for bubbles of other gas mixtures:}
As a blind test one should also measure the pH as a function of time for pure
argon bubbles. No pH change should occur.
Neither do we expect a pH change for oxygen-argon mixtures as the
main reaction product of the dissociated oxygen should be H$_2$O$_2$. 
\end{itemize}

\vspace{1cm}
\noindent
{\bf Acknowledgments:}
We are grateful to Michael Brenner for
our joint development of the Rayleigh-Plesset SL bubble approach.
This ongoing collaboration has always been a pleasure to all of us.
We thank him for 
lots of discussions and exchange 
over the last
years.
Without him this work would not have been possible.
Many thanks also go to Todd Dupont,
Siegfried Grossmann,
Leo Kadanoff,
Blaine Johnston,
David Oxtoby, and Ken Suslick. 
We would also like to acknowledge
Brad Barber's remark that high partial pressure air bubbles
experimentally behave similar to low partial pressure argon bubbles,
see ref.\ \cite{bar95}. 
-- 
Support for this work by
the Deutsche Forschungsgemeinschaft (DFG) under grant SBF185-D8 is
acknowledged.

\vspace{0.5cm}

\noindent
 e-mail addresses:\\
lohse@stat.physik.uni-marburg.de\\
hilgenfeldt@stat.physik.uni-marburg.de\\



\begin{thebibliography}{10}

\bibitem{fre34}
H. Frenzel and H. Schultes, Z. Phys. Chem. {\bf 27B},  421  (1934).

\bibitem{har39}
E.~N. Harvey, J. Am. Chem. Soc. {\bf 61},  2392  (1939).

\bibitem{kut62}
H. Kuttruff, Acustica {\bf 12},  230  (1962).

\bibitem{mar85}
M.~A. Margulis, Ultrasonics  157  (1985).

\bibitem{bre95b}
C.~E. Brennen, {\em Cavitation and Bubble Dynamics} (Oxford University Press,
  Oxford, 1995).

\bibitem{gai90}
D.~F. Gaitan, Ph.D. thesis, The University of Mississippi, 1990;
D.~F. Gaitan, L.~A. Crum, R.~A. Roy, and C.~C. Church, J. Acoust. Soc. Am. {\bf
  91},  3166  (1992).

\bibitem{cru94}
L.~A. Crum, Physics Today {\bf 47},  22  (1994);
S.~J. Putterman, Scientific American {\bf 272},  32  (1995).

\bibitem{bar91}
B.~P. Barber and S.~J. Putterman, Nature (London) {\bf 352},  318  (1991).

\bibitem{bar92}
B.~P. Barber and S.~J. Putterman, Phys. Rev. Lett. {\bf 69},  3839  (1992).

\bibitem{hil92}
R. Hiller, S.~J. Putterman, and B.~P. Barber, Phys. Rev. Lett. {\bf 69},  1182
  (1992).

\bibitem{hil94}
R. Hiller, K. Weninger, S.~J. Putterman, and B.~P. Barber, Science {\bf 266},
  248  (1994).

\bibitem{hil95}
R. Hiller and S.~J. Putterman, Phys. Rev. Lett. {\bf 75},  3549  (1995), also
  see the erratum, Phys. Rev. Lett. 77, 2345 (1996).

\bibitem{bar95}
B.~P. Barber, K. Weninger, R. L\"ofstedt, and S.~J. Putterman, Phys. Rev. Lett.
  {\bf 74},  5276  (1995).

\bibitem{loe93}
R. L\"ofstedt, B.~P. Barber, and S.~J. Putterman, Phys. Fluids A {\bf 5},  2911
   (1993).

\bibitem{loe95}
R. L\"ofstedt, K. Weninger, S.~J. Putterman, and B.~P. Barber, Phys. Rev. E
  {\bf 51},  4400  (1995).

\bibitem{wen95}
K. Weninger {\it et~al.}, J. Phys. Chem. {\bf 99},  14195  (1995).

\bibitem{cru75}
L.~A. Crum, J. Acoust. Soc. Am. {\bf 57},  1363  (1975).

\bibitem{hil96}
S. Hilgenfeldt, D. Lohse, and M.~P. Brenner, Phys. Fluids {\bf 8},  2808
  (1996).

\bibitem{ell69}
A. Eller, J. Acoust. Soc. Am. {\bf 46},  1246  (1969);
A. Eller and L.A. Crum, J. Acoust. Soc. Am. Suppl. {\bf 47},  762  (1970).

\bibitem{fyr94}
M.~M. Fyrillas and A.~J. Szeri, J. Fluid Mech. {\bf 277},  381  (1994).

\bibitem{bre96}
M.P. Brenner, D. Lohse, D. Oxtoby, and T.F. Dupont, Phys. Rev. Lett. {\bf 76},
  1158  (1996).

\bibitem{hol96}
G. Holt and F. Gaitan, Phys. Rev. Lett. {\bf 77},  3791  (1996).

\bibitem{ple54}
M. Plesset, J. Appl. Phys. {\bf 25},  96  (1954);
H.~W. Strube, Acustica {\bf 25},  289  (1971).

\bibitem{pro77}
A. Prosperetti, Quart. Appl. Math. {\bf 34},  339  (1977).

\bibitem{bre95}
M.P. Brenner, D. Lohse, and T.F. Dupont, Phys. Rev. Lett. {\bf 75},  954
  (1995).

\bibitem{holzfuss}
J.\ Holzfuss, priv. communication (1997). Similar experiments were
also made by F. Gaitan and G. Holt. The features of unstable SBSL
are also seen in the acoustic signal, see 
J. Holzfuss, M. R\"uggeberg, and A. Billo,
``Stosswellenentstehung und 
        Einzelblasen\-sonolumineszenz'',  
        Fortschritte der Akustik - DAGA 97, Bad Honnef:
        DPG GmbH (1997). 


\bibitem{hol94}
R.~G. Holt, D.~F. Gaitan, A.~A. Atchley, and J. Holzfuss, Phys. Rev. Lett. {\bf
  72},  1376  (1994).

\bibitem{bre96b}
M.~P. Brenner, S. Hilgenfeldt, and D. Lohse,  in {\em Nonlinear Physics of
  Complex Systems -- Current Status and Future Trends}, edited by J. Parisi,
  S.~C. M\"uller, and W. Zimmermann (Springer Lecture Notes in Physics, Berlin,
  1996), p.\ 79.

\bibitem{loh97}
D. Lohse, M. Brenner, T. Dupont, S. Hilgenfeldt, and B. Johnston,
Phys. Rev. Lett. {\bf 78},  1359  (1997).

\bibitem{wei53}
A. Weissler, J. Acoust. Soc. Am. {\bf 25},  651  (1953).

\bibitem{ver88}
R.~E. Verral and C.~M. Sehgal,  in {\em Sonoluminescence in Ultrasound: Its
  chemical, physical and biological effects}, edited by K. Suslick (VCH,
  Weinheim, 1988), p.\ 227.

\bibitem{suslick}
K. Suslick, Science 247, 1439 (1990);
E. B. Flint and 
K. Suslick, Science 253, 1397 (1991);
K. Suslick, E. B. Flint, M. W. Grinstaff, and K. A. Kemper,
J. Phys. Chem 97, 3098 (1993);
K. Suslick, ``The Chemistry of Ultrasound'', in The Yearbook of Science
and the Future 1994, Encyclopaedia Britannica, Chicago, pp 138-155;
T. Matula et al., Phys. Rev. Lett. 75, 2602 (1995).

\bibitem{sch36}
H. Schultes and H. Gohr, Angew. Chemie {\bf 49},  420  (1936).

\bibitem{temperaturefuss}
This value follows from the tabulated values for the enthalpy 
and the entropy changes, see e.g.\ G. M. Barrow,
``Physical Chemistry'', McGraw-Hill, New York, 1973.
It is the value for which the Gibbs free energy
$\Delta G$ passes through zero.


\bibitem{ber95}
L. Bernstein and M. Zakin, J. Phys. Chem. {\bf 99},  14619  (1995).

\bibitem{bar94}
B.~P. Barber {\it et~al.}, Phys. Rev. Lett. {\bf 72},  1380  (1994).

\bibitem{kon_priv}
L. Kondic, 1996, private communication.

\bibitem{har87}
E. Hart, Ch.~H. Fischer, and A. Henglein, J. Phys. Chem. {\bf 91},  4166
  (1987).

\bibitem{bre96d}
M.~P. Brenner, S. Hilgenfeldt, D. Lohse, and R. Rosales, Phys. Rev. Lett. {\bf
  77},  3467  (1996).

\bibitem{ray17}
Lord Rayleigh, Philos. Mag. {\bf 34},  94  (1917);
M. Plesset, J. Appl. Mech. {\bf 16},  277  (1949);
A. Prosperetti, J. Acoust. Soc. Am. {\bf 56},  878  (1974);
W. Lauterborn, J. Acoust. Soc. Am. {\bf 59},  283  (1976);
A. Prosperetti, Ultrasonics {\bf 22},  69  (1984).

\bibitem{pro77b}
A. Prosperetti, J. Acoust. Soc. Am {\bf 61},  17  (1977).

\bibitem{ple77}
M. Plesset and A. Prosperetti, Ann. Rev. Fluid Mech. {\bf 9},  145  (1977).

\bibitem{cru83}
L.~A. Crum, J. Acoust. Soc. Am. {\bf 73},  116  (1983).

\bibitem{pro88}
A. Prosperetti, L.A. Crum, and K.W. Commander, J. Acoust. Soc. Am. {\bf 83},
  502  (1988).

\bibitem{vuo96}
V.~Q. Vuong and A.~J. Szeri, Phys. Fluids {\bf 8},  2354  (1996).

\bibitem{kam93}
V. Kamath, A. Prosperetti, and F.N. Egolfopoulos, J. Acoust. Soc. Am. {\bf 94},
   248  (1993).

\bibitem{yas95}
K. Yasui, J. Acoust. Soc. Am. {\bf 98},  2772  (1995).


\bibitem{wen97}
Note that  using 
unrealistic
``effective'' viscosities as large as seven times the real viscosity of
water and also ``effective''
surface tensions as done by Putterman's group
(see \cite{loe93,bar94,bar95} and 
K.~Weninger, B. Barber and S. Putterman, Phys. Rev. Lett. {\bf 78},  1799
  (1997))
 is neither  justified  nor necessary to fit the
  data. 


\bibitem{wu93}
C.~C. Wu and P.~H. Roberts, Phys. Rev. Lett. {\bf 70},  3424  (1993);
Proc. R. Soc. London A {\bf 445},  323  (1994);
W.Moss, D. Clarke, J. White, and D. Young, Phys. Fluids {\bf 6},  2979
 (1994);
Phys. Lett. A {\bf 211},  69
  (1995);
  L. Kondic, J. Gersten, and C. Yuan, Phys. Rev. E 52, 4976 (1995);
  V. Q. Vuong and A. J. Szeri, Phys.\ Fluids 8, 2354 (1996). 
  

\bibitem{hil97}
S. Hilgenfeldt, M.~P. Brenner, S. Grossmann, and D. Lohse, submitted to J.
  Fluid Mech. (1997).

\bibitem{ber96}
L. Bernstein, M. Zakin, E. Flint, and K. Suslick, J. Phys. Chem. {\bf 100},
  6612  (1996).

\bibitem{lid91}
D.~R.~Lide (editor), {\em Handbook of Chemistry and Physics} (CRC Press, Boca
  Raton, 1991).

\bibitem{lan69}
Landolt and B\"ornstein, {\em Zahlenwerte und Funktionen aus Physik und Chemie}
  (Springer, Berlin, 1969).

\bibitem{mat97}
T.J. Matula, S.~M. Cordry, R.~A. Roy, and L.~A. Crum, J. Acoust. Soc. Am.,
submitted (1997).

\bibitem{cho96b}
D. Chow, K. Weninger, B.~P. Barber, and S.~J. Putterman, J. Acoust. Soc. Am.
  {\bf 100},  2718  (1996).

\end{thebibliography}

\end{document}